\documentclass[conference]{acmsiggraph}

\usepackage{amsmath,amsfonts}
\usepackage{subcaption}
\usepackage{wrapfig}
\usepackage{array}
\usepackage{stfloats}
\usepackage{balance}
\newcolumntype{C}[1]{>{\centering\arraybackslash}p{#1}}
\newcommand\Tstrut{\rule{0pt}{2.6ex}}
\newcommand\Bstrut{\rule[-0.9ex]{0pt}{0pt}}

\TOGonlineid{}
\TOGvolume{0}
\TOGnumber{0}
\TOGarticleDOI{}
\TOGprojectURL{}
\TOGvideoURL{}
\TOGdataURL{}
\TOGcodeURL{}

\title{Geodesics using Waves: Computing Distances using Wave Propagation}

\author{Ayushi Sinha\thanks{e-mail:asinha8@jhu.edu} \qquad Michael Kazhdan\thanks{email:misha@cs.jhu.edu} \\Department of Computer Science, the Johns Hopkins University, Baltimore, MD}
\pdfauthor{Ayushi Sinha}


\begin{document}

 \teaser{
   \includegraphics[width=0.91\textwidth]{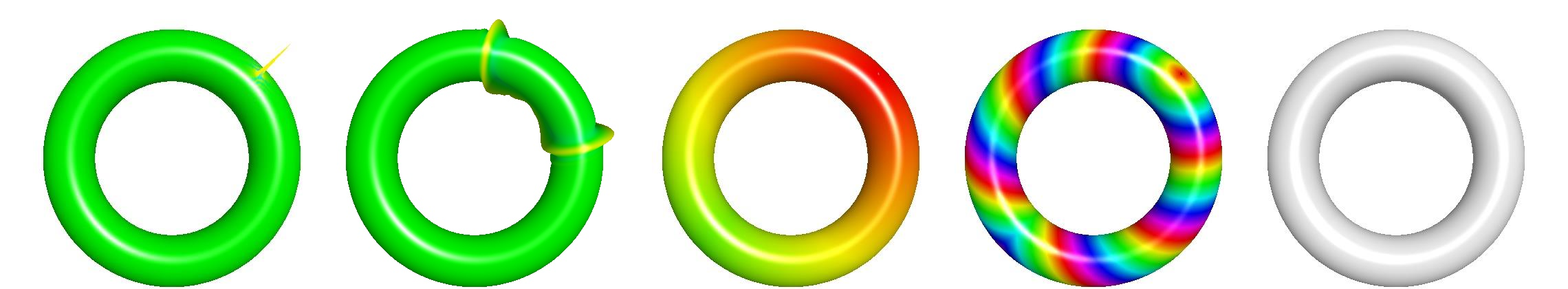}
   \caption{Approximate geodesic distances using wave propagation: (a) Wave propagation is initiated by setting the initial signal such that its only peak is at the source of the wave; (b) repeatedly solving the wave equation on this signal propagates the wave outward from the source, across the shape; here yellow indicates the positive part of the wavefront, and green indicates that the height of the wave is approximately zero; (c) the wave propagation is used to define a pseudo-distance function, (d) which, in turn, is used to approximate geodesic distances; (e) the error compared to exact geodesics, where 0 error is indicated by white.}
   \label{fig:teaser}
 }

\maketitle

\begin{abstract}

In this paper, we present a new method for computing approximate geodesic distances. We introduce the wave method for approximating geodesic distances from a point on a manifold mesh. Our method involves the solution of two linear systems of equations. One system of equations is solved repeatedly to propagate the wave on the entire mesh, and one system is solved once after wave propagation is complete in order to compute the approximate geodesic distances up to an additive constant. However, these systems need to be pre-factored only once, and can be solved efficiently at each iteration. All of our tests required approximately between $300$ and $400$ iterations, which were completed in a few seconds. Therefore, this method can approximate geodesic distances quickly, and the approximation is highly accurate.


\end{abstract}

\begin{CRcatlist}
  \CRcat{I.3.5}{Computer Graphics}{Computational Geometry and Object Modeling}{Geometric algorithms, languages, and systems}
\end{CRcatlist}

\keywordlist




\copyrightspace

\section{Introduction}

When a pebble is dropped in a body of water, the impact causes waves to form and propagate in all directions moving away from the point of impact. If the medium is uniform, then the waves will travel uniformly in every direction, and therefore reach points that are equidistant from the source of the wave at the same time. Similarly, waves can also be propagated on manifold meshes, and expected to reach points equidistant from the source of the wave at approximately the same time. Therefore, we can define a function that measures the time required for a wave to travel from the source of the wave, $x$, to any other point, $y$, on the mesh.This function can be thought of as a pseudo-distance function, since it does not tell us what the real distance between $x$ and $y$ is, but it gives us some sense of how far the two points are from each other.

This pseudo-distance function can be used to approximate exact geodesic distances. We compute the normalized gradient of the pseudo-distance function for every face in the mesh, and then solve a Poisson equation to compute a function which has the same gradient as the normalized gradient of the pseudo-distance function. This produces the approximate geodesic distances we want, up to an additive constant \cite{Crane_et_al.2013}.

\section{Related Work}

Solving the special case of the Eikonal equation 
\begin{equation}
	|\nabla\phi| = 1
	\label{eq:eikonal}
\end{equation}
subject to boundary conditions $\phi|_{\partial \Omega} = 0$, produces the shortest distance from the boundary, $\partial\Omega$, to any point in $\Omega$, or the geodesic distance. However, this partial differential equation is non-linear and hard to solve without using some type of iterative relaxation approach~\cite{Hysing_and_Turek2005}. The fast marching method \cite{Kimmel_and_Sethian1998} and fast sweeping method \cite{Xu_et_al.2010} on triangulated manifolds are two popular algorithms for computing geodesic distances using this approach. \cite{Mitchell_et_al.1987} developed another method to compute the exact piecewise linear geodesic distance between a source and all other vertices of a manifold mesh. However, this algorithm was not implemented until much later by \cite{Surazhsky_et_al.2005}. They also extended the work of \cite{Mitchell_et_al.1987} with an approximation algorithm with bounded errors and faster runtime. 

One of the most recent methods for approximating geodesics is the heat method \cite{Crane_et_al.2013}, which offers several improvements over previous methods. One of the advantages it offers is that it is simple to implement since it requires the solution of two standard linear systems of equations which can be solved efficiently, making this method very fast. Our method is similar to this method in that our method also requires the solution of two linear systems of equations, which can be prefactored~\cite{Chen_et_al.2008}, and then solved efficiently. However, our method requires one of these systems of equations to be solved repeatedly.

\section{The Wave Method}



Let $\mathcal{M = (V,E,F)}$ be a manifold mesh with $\mathcal{V}$ vertices, $\mathcal{E}$ edges, and $\mathcal{F}$ faces. Then, the wave equation defined over $\mathcal{M}$ can be written as 
\begin{align} \label{eq:wave}
	\phi_t : \mathcal{M} &\times \mathbb{R} \rightarrow \mathbb{R} \nonumber\\
	\frac{ \partial^2 } { \partial t^2 } \phi_t &= \mu \Delta \phi_t, \\
	\phi_0 &= \phi \nonumber\\
	\frac{ \partial } { \partial t} \phi_0 &= 0 \nonumber \\
	\mu &= 1 \nonumber 
\end{align}
where $\Delta$ is the Laplace operator, and $\phi$ is the initial signal. We will talk about how we set the initial signal in the next section. We solve this equation to propagate a wave over time on the surface of $\mathcal{M}$.

By tracking this wave propagation, we can define a pseudo-distance function over the manifold. We can then compute the normalized gradient, $G$, of the pseudo-distance function, and solve a Poisson equation with fixed boundary to approximate a function with the same gradient as $G$. Since, the gradient of the geodesic distance function has unit length, the solution of the Poisson equation gives us the approximate geodesic distances over the manifold up to an additive scale \cite{Crane_et_al.2013}.

\subsection{Wave Propagation}
We start the wave propagation at a source vertex, $v$, by setting an initial signal, $\phi_0$, such that its only peak is at $v$. Then, we flow Eq.~\ref{eq:wave} repeatedly, and track the leading wavefront until we have recorded a pseudo-distance at each vertex in the mesh. We will describe our initial signal and wave propagation in this section.

\subsubsection{Initial Value}
The initial signal is set so that the signal is high only at the vertex that is chosen to be the source of the wave. The signal at all other vertices is set to 0, while at the source vertex it can be set to any value greater than zero. We set the initial signal such that for each vertex, $v \in \mathcal{V}$,

\begin{equation*}
	 \phi_0(v) =
	  \begin{cases}
		    1, & \text{ if $v$ is the source of the wave,} \\
		    0, & \text{ otherwise. }
	  \end{cases}
\end{equation*}

Since, the change in this signal at time $t=0$ is defined to be $0$ (Eq.~\ref{eq:wave}), the signal in the time-step before the first time-step is defined to be the same as $\phi_0$.

\subsubsection{Implicit Formulation}
We discretize the flow using the following implicit formulation for a mesh with $n$ vertices:
\begin{equation*}
	\left\langle \frac{ \phi_{t+\delta} - 2\phi_t + \phi_{t-\delta} }{ \delta^2 } , b_v \right\rangle = \left\langle \Delta \phi_{t+\delta} , b_v \right\rangle,
\end{equation*}
where $b_v, v=1 \ldots n$ is the hat-basis function at vertex $v \in \mathcal{V}$, and $\delta$ is the time-step. By expressing $\phi_t$ as a linear combination of these basis functions,
\begin{equation*}
	\phi_t(p) = \sum_{w \in V} \xi_w(t) b_w(p) ,
\end{equation*}
we can write
\begin{align}  \label{eq:semi-imp_wave_incomp}
	&\frac{1}{\delta^2} \sum_{w \in V} \left\langle\xi_w(t+\delta) b_w - 2\xi_w(t) b_w + \xi_w(t-\delta) b_w , b_v \right\rangle \nonumber \\
	&\hspace{30mm}= \sum_{w \in V} \left\langle \Delta \xi_w(t+\delta) b_w , b_v \right\rangle, \forall v \in V \nonumber \\
	\Rightarrow &\frac{1}{\delta^2} \sum_{w \in V} \big[ \xi_w(t+\delta) - 2\xi_w(t) + \xi_w(t-\delta) \big] \left\langle b_w , b_v \right\rangle \nonumber \\
	&\hspace{30mm}= \sum_{w \in V} \xi_w(t+\delta) \left\langle \Delta b_w , b_v \right\rangle, \forall v \in V \nonumber \\
\end{align}

Since, $M = \left\langle b_w , b_v \right\rangle$ is the mass matrix, and  $S = -\left\langle \Delta b_w , b_v \right\rangle$ is the stiffness matrix, Eq.~\ref{eq:semi-imp_wave_incomp} becomes
\begin{align} \label{eq:semi-imp_wave}
	\big[ 2\vec{\xi}(t) - \vec{\xi}(t-\delta) \big] M = (\mu \delta^2 S + M ) \vec{\xi}(t+\delta) 
\end{align}

This equation is of the form $Ax=b$, where $x$ is unknown, and $A$ is a symmetric, positive definite matrix. 

\begin{figure}[!t]
  	\centering
  	\includegraphics[width=0.475\textwidth]{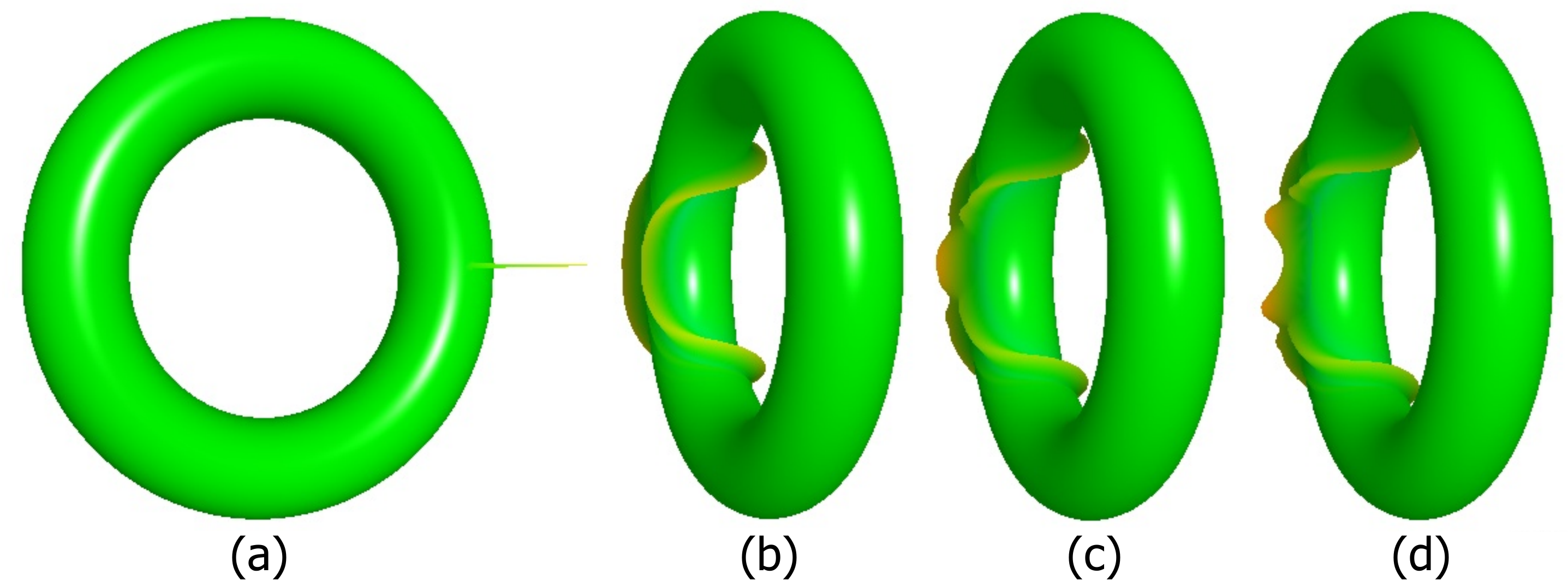}
\vspace{0.0001mm}
  	\caption{When two waves with equal amplitudes traveling towards each other collide, the amplitude of the wave at collision is the sum of the amplitudes of the two colliding waves. This is called constructive collision. We can see an instance of constructive collision on the torus model. (a) shows the source of the wave, and (b) shows the wavefront just before collision. Constructive collision resulting in a wave with a higher amplitude than the colliding wavefronts can be seen as the bump where the collision occurred in (c). (d) shows these bumps propagating past the point of collision.}
  	\label{fig:collision}
\end{figure}

The source of the wave and the wave propagation produced can be seen in Figures~\ref{fig:collision}a and~\ref{fig:collision}b, respectively. We want to be able to track the leading tall {\em wavefront} traveling from the source across the shape in order to define a pseudo-distance function on $\mathcal{M}$.

\subsection{Pseudo-Distance Function}
\subsubsection{Naive Approach}
Since we expect the largest amplitude of the wave to be at the wavefront, we first tried to track it by tracking its peak. That is, for each vertex $i$ on the manifold mesh, we record the time at which the wave attains its maximum value. This distance function, $d$, can be formulated as
\begin{align*}
	d:& \mathcal{V} \rightarrow \mathbb{R} \\
	d(v) = &\operatorname{arg\,max}_t \phi_t(v).
\end{align*}
This distance formulation, however, runs into problems at points of collision. When there is a constructive collision, the amplitude of the wave at the point of collision is the sum of the amplitudes of the two colliding waves. Therefore, at the point of collision, the amplitude of the wave becomes larger than the amplitude of the wavefront  (Fig.~\ref{fig:collision}c,d), causing artifacts in the distance function.

\begin{figure}[!t]
  \centering
  \includegraphics[width=2.85in]{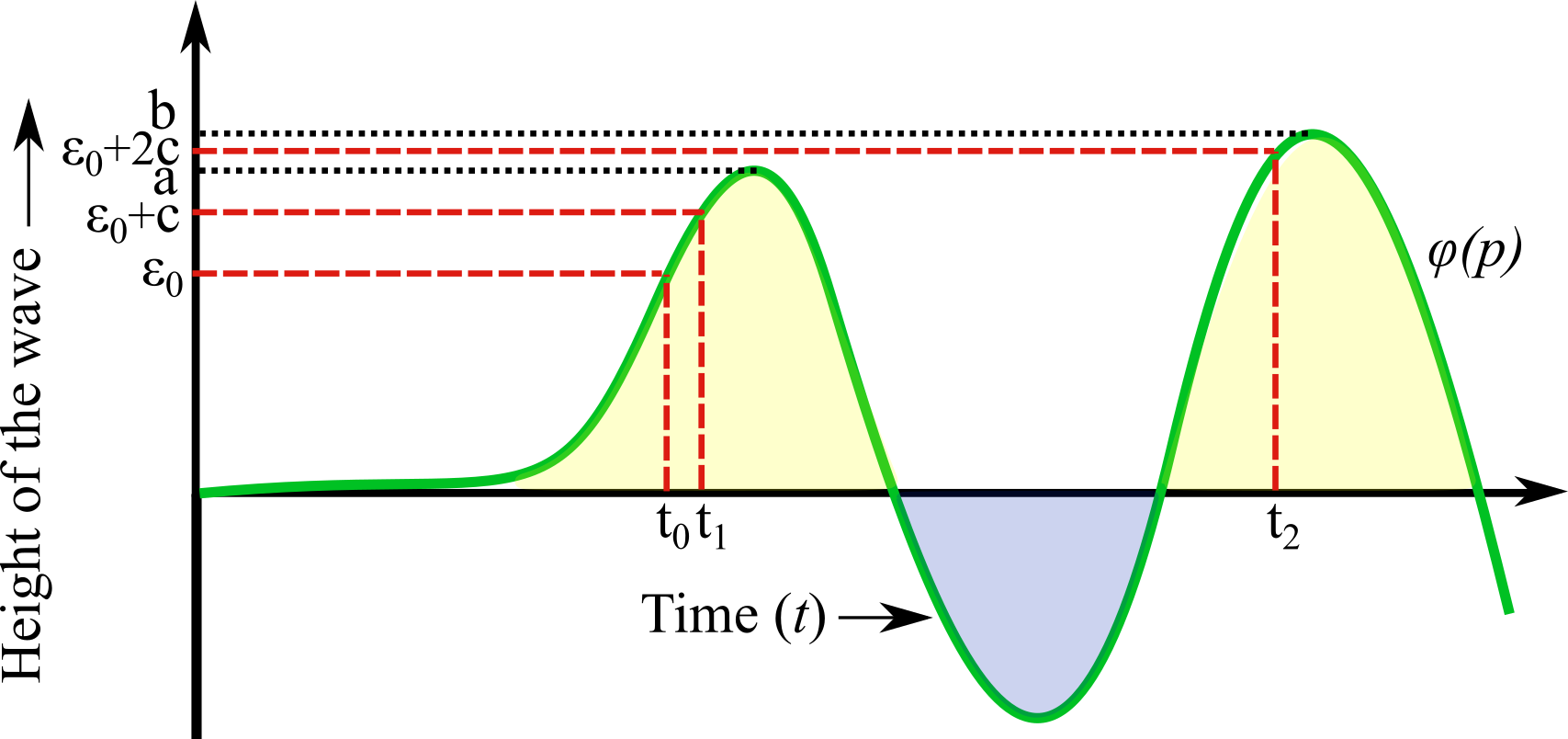}
  \caption{Instability of the distance function $d(p)$: Given that at $\epsilon = \epsilon_0$, $d(p) = t_0$, adding a small constant $c$ to $\epsilon_0$, such that $\epsilon_0+c < a$, where $a$ is the amplitude of the first wave in the figure, results in a small change in the distance function, $d(p) = t_1$, which is close to $t_0$. However, if $c$ is added to $\epsilon_0$ again, such that $\epsilon_0+2c > a$, then the change in the distance function is extremely large, since $d(p) = t_2$ is a lot larger than $t_1$. This error can happen if $\epsilon$ is larger than the height of the wavefront, and there is a larger wave following this wavefront due to a collision.}
  \label{fig:instability}
\end{figure}

\begin{figure*}[!b]
  	\begin{subfigure}{0.5\textwidth}
		\includegraphics[width=\textwidth]{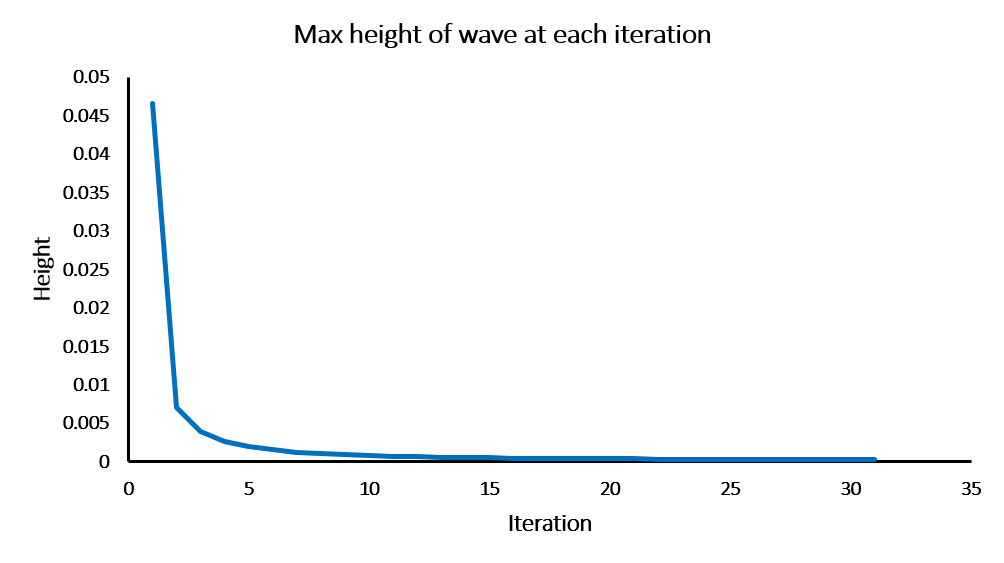}
		\caption{}
	\end{subfigure}
	\begin{subfigure}{0.5\textwidth}
		\includegraphics[width=\textwidth]{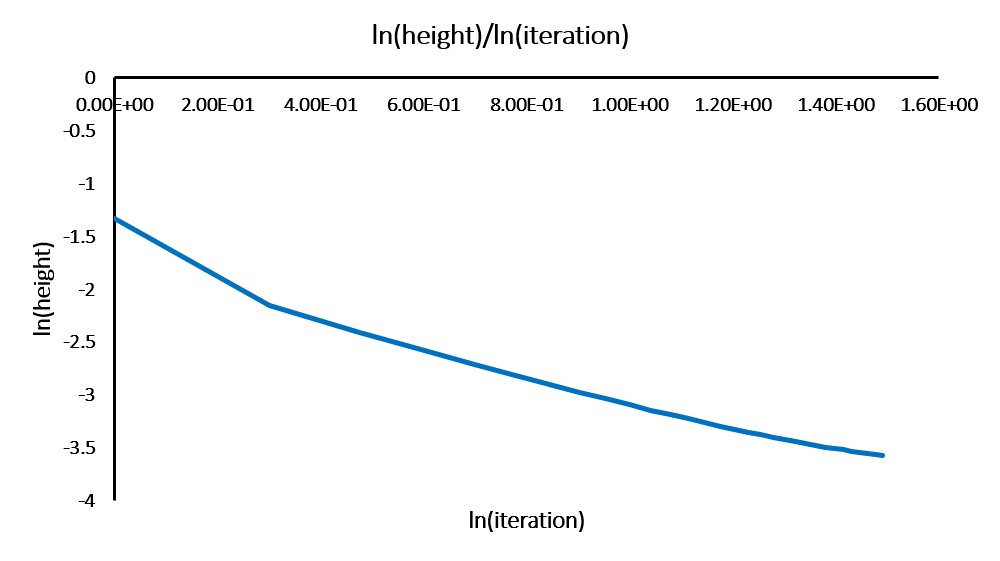}
		\caption{}
	\end{subfigure}
	\caption{Wave propagated on a sphere model with $\delta = 0.05$. (a) Maximum height of the wave plotted per iteration; (b) the plot for log(height)/log(iteration) has average slope of ~$-3$.}
	\label{ht1}
\end{figure*}

\subsubsection{Our Approach}
Instead, we track the time at which the wave first attains a prescribed height. That is, for each vertex $v$ on the mesh, we record the time at which the wave first achieves a small positive value, $\epsilon$.
\begin{align}
	d: \mathcal{V}& \rightarrow \mathbb{R} \nonumber \\
	d(v, \epsilon) = inf&\{t: \phi_t(v)=\epsilon\} \label{eq:track_start}.
\end{align}

This distance formulation records the time at which the wavefront reaches a vertex $v$. This formulation is able to avoid the problems due to collision that arose in the previous formulation. However, this formulation is unstable.

\subsubsection{Instability}
As the wavefront propagates across the shape, its amplitude decreases gradually as the energy of the wave dissipates to larger areas and over more vertices. We have to take this into consideration when choosing a value for $\epsilon$, because if the height of the wavefront is smaller than $\epsilon$ at a vertex $v$, then no $d(v,\epsilon)$ value will be recorded for the wavefront. However, a $d(v,\epsilon)$ can be recorded at a later time due to a larger wave resulting from a collision, for instance. In other words, a slight change in $\epsilon$ can result in a big change in $d$ (Fig.~\ref{fig:instability}), making it important for $\epsilon$ to remain below the expected height of the wavefront throughout the wave propagation. Therefore, we need to choose $\epsilon$ conservatively, while also making sure to not choose too small a value, since this will make $d$ sensitive to noise. 

\subsubsection{Spatially Varying $\epsilon$}
In order to address this problem, we let $\epsilon$ vary as we move away from the source. The idea is to estimate the height  of the wavefront, $h(i)$, at every iteration, $i$, of the wave propagation. Using this information, we can estimate a function for $\epsilon$, $\epsilon(i)$, which is smaller than $h(i)$ at every iteration. In order to do this, we recorded the maximum height of the wave with $\delta = 0.05$ at each iteration on a sphere model to avoid peaks resulting from collision (Fig.~\ref{ht1}a):
\begin{equation*}
	h(i) = \operatorname{arg\,max}_v \phi(v) \textrm{, for } i=1,2,\ldots
\end{equation*}

Empirically, we observed that the height of the wave was approximately inversely proportional to the iteration of wave propagation. From the graph, we expect $h(i) \approx ci^a$, where $i$ is the iteration, $c$ is a constant, and $a$ is a negative number. Assuming $c=1$, since it is easier to estimate than $a$, we plot $\ln{h(i)}$ against $\ln{i}$ (Fig.~\ref{ht1}b) to estimate the slope, $a$, of the graph: 
\begin{equation*}
\begin{split}
	h(i) = i^a &\Rightarrow \ln{h(i)} = a \ln{i} \\
	&\Rightarrow \frac{\ln{h(i)}}{\ln{i}} = a 
\end{split}
\end{equation*}

We found the average slope for $\delta = 0.05$ to be approximately $-3$. Therefore,
\begin{equation*}
	h(i) = \frac{c}{i^3}
\end{equation*}

\begin{figure*}[!t]
  	\begin{subfigure}{0.5\textwidth}
		\includegraphics[width=\textwidth]{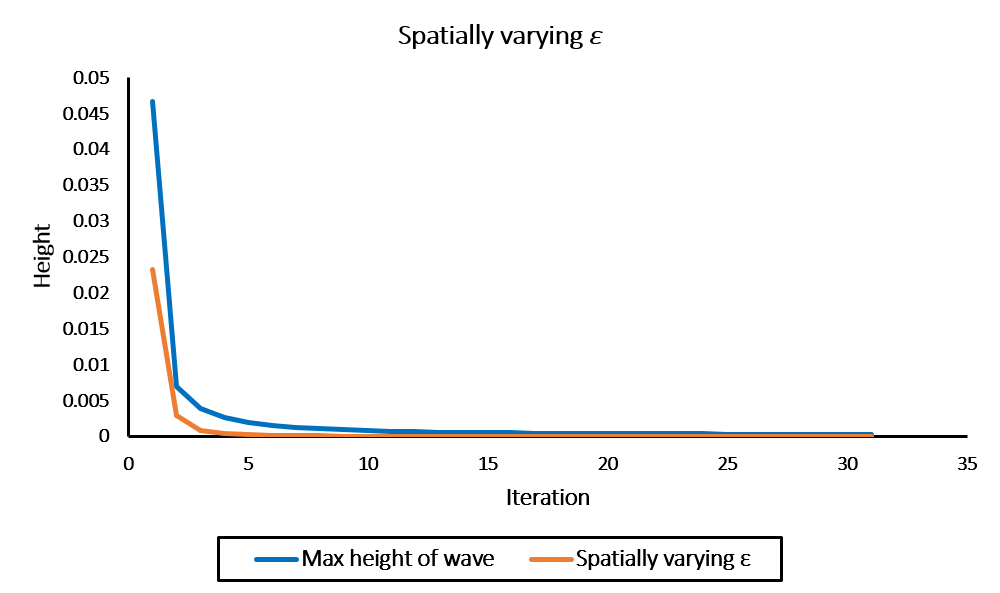}
		\caption{}
	\end{subfigure}
	\begin{subfigure}{0.5\textwidth}
		\includegraphics[width=\textwidth]{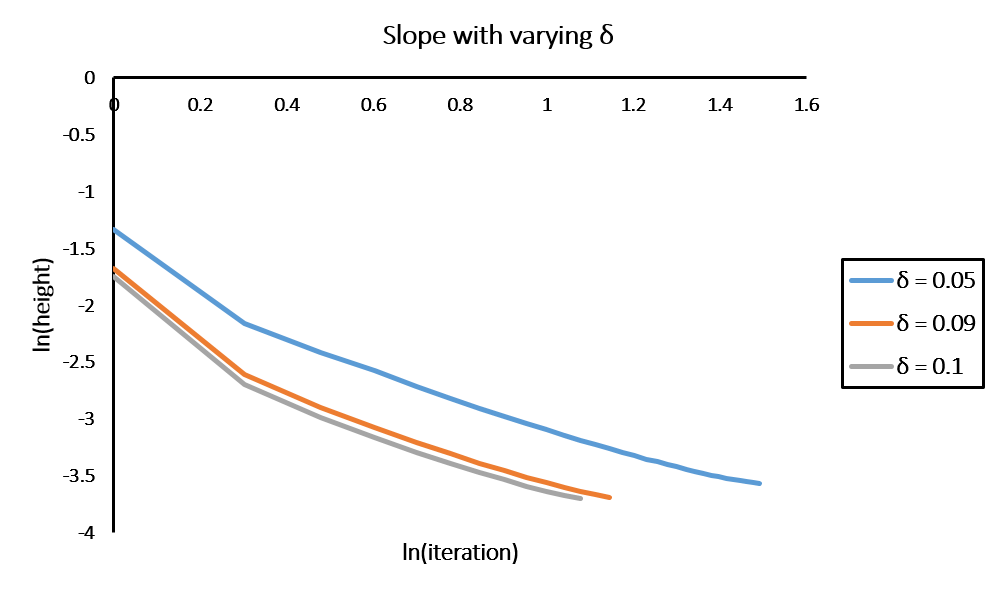}
		\caption{}
	\end{subfigure}
	\caption{(a) Spatially varying $\epsilon$ plotted per iteration along with the maximum height of the wave; (b) plot for log(height)/log(iteration) for varying $\delta$.}
	\label{ht2}
\end{figure*}

Now, we can compute the constant $c$. We want $\epsilon(i)$ to vary in the same manner as $h(i)$. At the same time, we also want $\epsilon(i)$ to remain lower than $h(i)$ at every iteration. Therefore, we set $c$ to be equal to half the maximum height of the wave at the first iteration (Fig.~\ref{ht2}a).
\begin{equation*}
\begin{split}
	\epsilon(i) &= \frac{h(1)}{2 i^{3} }
\end{split}
\end{equation*}

We also observed wave propagation with varying $\delta$ to understand how the height of the wave changes as it propagates, and found the value of $a$ to vary with $\delta$. As $\delta$ increases, the rate at which the height of the wave falls per iteration also increases. Therefore, $a$ decreases as $\delta$ increases (Fig.~\ref{ht2}b).

\subsubsection{Time Estimation}
Now that we have a formulation for $\epsilon$ for every iteration, we need to compute $d(v,\epsilon)$ at all the vertices as shown in Eq.~\ref{eq:track_start}. That is, we need to record the time at which $\phi_t(v)$ becomes exactly equal to the $\epsilon$ at that iteration. Since absolute equality is improbable, we need to interpolate to estimate when the wave achieved height $=\epsilon$. We linearly interpolate between the time $t1$ at which height $<\epsilon$ was achieved and $t2$ at which height $>\epsilon$ was achieved to estimate the time $t$ at which height $=\epsilon$.

\subsection{Geodesic Distance Function}
Once we have a pseudo-distance function defined at every vertex, we can compute the normalized gradient, $G$, of this function, and use it to find a function with the same gradient as $G$ by solving a Poisson equation as in \cite{Crane_et_al.2013}. 

Since the gradient of the true distance function has unit length, the function we obtain by solving the Poisson equation is an approximation to the geodesic distance function we want, up to an additive constant.

\subsubsection{Gradient}
For a manifold $\mathcal{M = (V,E,F)}$, we can compute the gradient, $G$, of our pseudo-distance function defined on the vertices ($\mathcal{V}$) of our mesh: 
\begin{equation*}
\begin{split}
	G : \mathbb{R}^{|\mathcal{V}|} \rightarrow &\mathbb{R}^{2|\mathcal{T}|}\\
\end{split}
\end{equation*}
We compute normalized $G$, $\bar{G}$, per mesh face $\mathcal{F}$, and compute its divergence. We can compute two gradient fields per mesh face:
\begin{equation}
\begin{split}
	\vec{w_1} &= p(v_1 - v_0) + q(v_2-v_0) \\
	\vec{w_2} &= s(v_1 - v_0) + t(v_2-v_0)
	\label{eq:gradient1}
\end{split}
\end{equation}
where $p, q, s$ and $t$ are unknown. Then,
\begin{equation*}
\begin{split}
	\langle \vec{w_1} , \vec{w_2} \rangle &= \langle p(v_1 - v_0) + q(v_2-v_0) , s(v_1 - v_0)+t(v_2 - v_0) \rangle \\
	& = \begin{pmatrix}
	  s \hspace{2mm} t
	 \end{pmatrix}
	 K
	 \begin{pmatrix}
	  p  \\
	  q
	 \end{pmatrix}
\end{split}
\end{equation*}
where $K_{ij} = \langle v_i - v_0, v_j-v_0\rangle$, $i,j=1,2$. 

We also know by definition that the gradients satisfy
\begin{equation}
\begin{split}
	\langle \vec{w_1} , v_1 - v_0 \rangle &= d_1 - d_0 \\
	\langle \vec{w_1} , v_2 - v_0 \rangle &= d_2 - d_0 
	\label{eq:gradient2}
\end{split}
\end{equation}
\begin{equation}
\begin{split}
	\langle \vec{w_2} , v_1 - v_0 \rangle &= d_1 - d_0 \\
	\langle \vec{w_2} , v_2 - v_0 \rangle &= d_2 - d_0 
\end{split}
\end{equation}
where $d_v$ are values of the pseudo-distance function at each vertex. We can rewrite Eq.~\ref{eq:gradient2} as
\begin{equation*}
\begin{split}
	\langle p(v_1 - v_0) + q(v_2-v_0) , v_1 - v_0 \rangle &= d_1 - d_0, \\
	\langle p(v_1 - v_0) + q(v_2-v_0) , v_2 - v_0 \rangle &= d_2 - d_0. \\
\end{split}
\end{equation*}
\begin{figure}[!b]
  \centering
  	\begin{subfigure}{38mm}
		\includegraphics[scale=0.22]{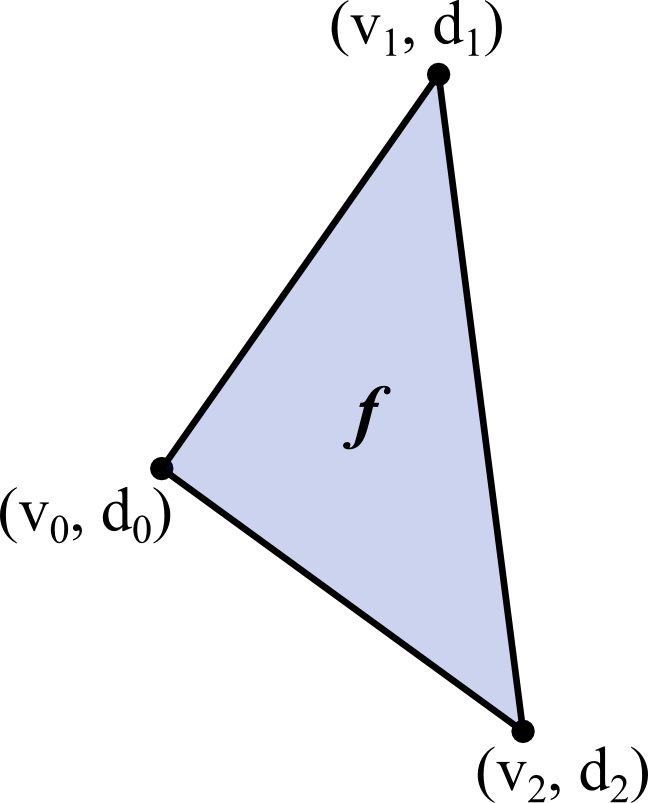}
		\caption{}
	\end{subfigure}
	\begin{subfigure}{30mm}
		\includegraphics[scale=0.18]{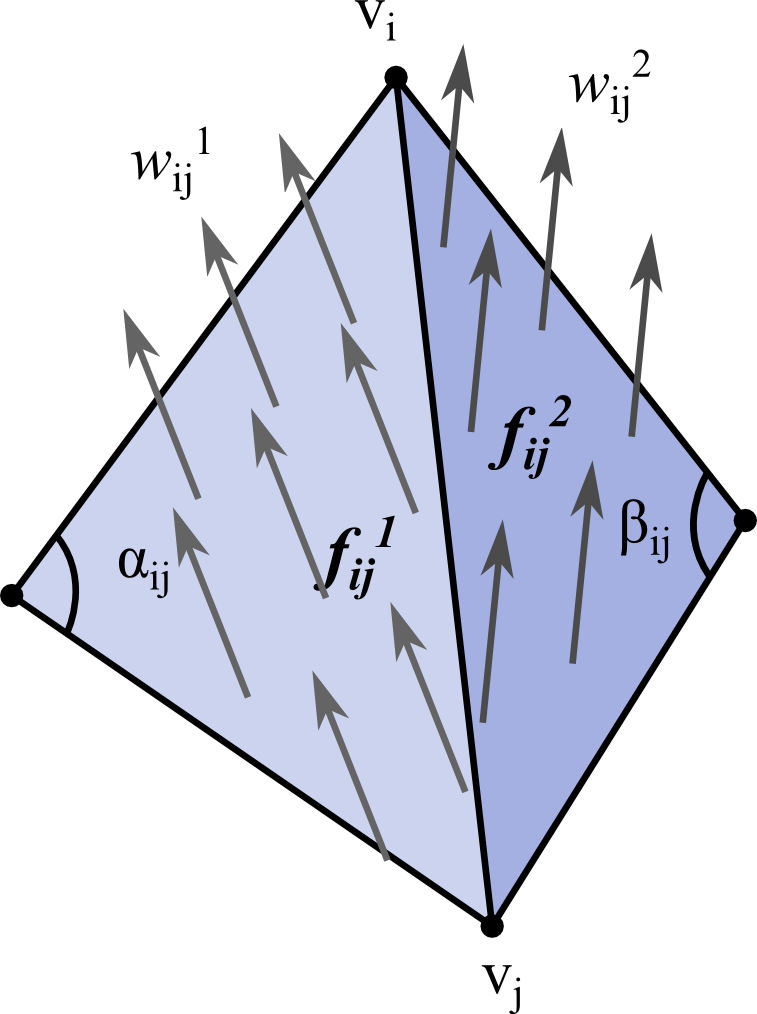}
		\caption{}
	\end{subfigure}
\vspace{-1mm}
  \caption{}
\vspace{-4mm}
  \label{fig:tri}
\end{figure}
We have two equations and two unknowns, $p$ and $q$, which can be computed by solving
\[
	K
	 \begin{pmatrix}
	  p  \\
	  q
	 \end{pmatrix} = 
	 \begin{pmatrix}
	  d_1-d_0  \\
	  d_2-d_0
	 \end{pmatrix}
\]
Similarly, we can compute $s$ and $t$, which allows us to compute the normalized Eq.~\ref{eq:gradient1} per mesh face. 
Then, we define a difference function at each edge in the mesh by computing the mean gradient at the two triangles incident upon each edge:
\begin{equation}
\begin{split}
	\widetilde{G} : \mathbb{R}^{|\mathcal{T}|} \rightarrow & \mathbb{R}^{|\mathcal{E}|}\\ 
	e = (v_i,v_j) \rightarrow \frac{1}{2} \langle v_j - v_i &, w_{ij}^1 + w_{ij}^2 \rangle, 
\label{eq:edge_grad}
\end{split}
\end{equation}
where $\mathcal{E}$ is the set of edges.

\begin{table*} [!t]
\caption{This table shows a numerical comparison between the wave method, heat method, and the exact method.}
\resizebox{1\textwidth}{!}{ 
	\begin{tabular}{ c|c||c|c|c|c|c||c|c|c|c||c} 
		\multicolumn{1}{c|}{MODEL} 
		& \multicolumn{1}{c||} {FACES}
		& \multicolumn{5}{c||} {WAVE METHOD} 
		& \multicolumn{4}{c||} {HEAT METHOD}
		& \multicolumn{1}{c} {EXACT} \Tstrut \Bstrut \\
	\hline
		\multicolumn{1}{c|}{}
		 & \multicolumn{1}{c||}{}
		 & \multicolumn{1}{C{0.5cm}|}{$\delta$}
		 & \multicolumn{1}{C{1.5cm}|}{MEAN RAW ERROR}
		 & \multicolumn{1}{C{2cm}|}{MEAN RELATIVE ERROR}
		 & \multicolumn{1}{C{1.5cm}|}{MAX RAW ERROR}
		 & \multicolumn{1}{C{1cm}||}{TIME (s)}
		 & \multicolumn{1}{C{1.5cm}|}{MEAN RAW ERROR}
		 & \multicolumn{1}{C{2cm}|}{MEAN RELATIVE ERROR}
		 & \multicolumn{1}{C{1.5cm}|}{MAX RAW ERROR}
		 & \multicolumn{1}{C{1cm}||}{TIME (s)}
		 & \multicolumn{1}{C{1cm}}{TIME (s)} \Tstrut \Bstrut \\
	\hline 
	\hline
		BUNNY & 28k & 0.005 & 0.00574625 & 0.00500049 & 0.058528 & 2.5 & 0.0137430 & 0.0112878 & 0.0974171 & 0.205 & 0.573 \Tstrut \Bstrut \\ 
		HORSE & 96k & 0.006 & 0.0156859 & 0.0131532 & 0.0560383 & 6.288 & 0.00470463 & 0.00425243 & 0.0312462 & 0.859 & 4.016 \\ 
		KITTEN & 106k & 0.005 & 0.00684904 & 0.00505089 & 0.0367825 & 8.101 & 0.00478574 & 0.00453304 & 0.0197093 & 1.477 & 5.335 \\ 
		BIMBA & 149k & 0.004 & 0.00693687 & 0.00742395 & 0.06061 & 12.579 & 0.00773516 & 0.00741442 & 0.0512189 & 2.476 & 7.641 \\ 
		RAMESSES & 1.6M & 0.09 & 0.0245474 & 0.0255807 & 0.131039 & 7.175 & 0.0111101 & 0.00909884 & 0.237803 & 36.227 & 37.638 \\ 
	\hline
	\end{tabular}
	} 
	\label{table:error} 
\end{table*}

\subsubsection{Poisson Equation}
Once we have gradients defined at each edge, we solve the Poisson equation to obtain the geodesic distance function on the mesh. Let $L$ be the Laplacian which can be computed as
\begin{equation*}
	  L_{ij} \in \mathbb{R}^{|\mathcal{V}|\times|\mathcal{V}|}=
	  \begin{cases}
		    \frac{\cot{\alpha_{ij}}+ \cot{\beta_{ij}}}{2}, & \text{ if } j \in N(i)\\
		    -\sum_{k\in N(i)} L_{ik}, & \text{ if } i = j\\
		    0, & \text{ otherwise, }
	  \end{cases}
\end{equation*}
\begin{table} [!b]
\caption{This table shows runtime comparison between the wave method and the exact method for increasing number of faces.}
\resizebox{0.5\textwidth}{!}{ 
	\begin{tabular}{ c||c|c|c|c||c }
		   \multicolumn{1}{c||} {FACES}
		& \multicolumn{4}{c||} {WAVE METHOD} 
		& \multicolumn{1}{c} {EXACT} \Tstrut \Bstrut \\
	\hline
		\multicolumn{1}{c||}{}
		 & \multicolumn{1}{C{1.5cm}|}{MEAN RAW ERROR}
		 & \multicolumn{1}{C{2cm}|}{MEAN RELATIVE ERROR}
		 & \multicolumn{1}{C{1.5cm}|}{MAX ERROR}
		 & \multicolumn{1}{C{1cm}||}{TIME (s)}
		 & \multicolumn{1}{C{1cm}}{TIME (s)} \Tstrut \Bstrut \\
	\hline 
	\hline
		20k & 0.016717 & 0.026019 & 0.051859 & 2.031 & 0.388 \Tstrut \Bstrut \\
		80k & 0.012966 & 0.027448 & 0.040692 & 5.32 & 3.682 \\
		320k & 0.007401 & 0.01482 & 0.020993 & 18.198 & 44.233 \\
	\hline
	\end{tabular}
	} 
	\label{table:resolution} 
\end{table}
where $\alpha_{ij}$ and $\beta_{ij}$ are as shown in Fig.~\ref{fig:tri}b, and $N(i)$ is the 1-ring neighborhood of vertex $i$. $L$ can also be defined as 
\[
	L = D^t \Lambda D,
\]
where $D$ is the matrix representing a linear operator in bases $\{m_1, m_2, \ldots, m_n\}$ and $\{o_1, o_2, \ldots, o_n\}$ defined on real vector spaces $M$ and $O$, making $D^t$ the matrix representing the dual operator in the corresponding dual bases $\{m_1^*, m_2^*, \ldots, m_n^*\}$ and $\{o_1^*, o_2^*, \ldots, o_n^*\}$ in $M^*$ and $O^*$, and $\Lambda = \langle l,m \rangle$, $l,m \in M$ is the map from real vector space to dual space. We define $D$ and $\Lambda$ as:
\begin{equation*}
	  D_{ev} \in \mathbb{R}^{|\mathcal{E}|\times|\mathcal{V}|}=
	  \begin{cases}
		    1, & \text{ if vertex $v$ is the end point of edge $e$} \\
		    -1, & \text{ if vertex $v$ is the starting point of $e$} \\
		    0, & \text{ otherwise, }
	  \end{cases}
\end{equation*}
and
\begin{equation*}
	  \Lambda_{ef} \in \mathbb{R}^{|\mathcal{E}|\times|\mathcal{E}|}=
	  \begin{cases}
		    \frac{\cot{\alpha_{ij}}+ \cot{\beta_{ij}}}{2}, & \text{ if } e=f\\
		    0, & \text{ otherwise. }
	  \end{cases}
\end{equation*}
We want to compute
\begin{equation*}
\begin{split}
	  w &= (D^t\Lambda D)^{-1} D^t\Lambda \tilde{G}\\
	  \Rightarrow \hspace{3mm} w &= L^{-1} D^t\Lambda \tilde{G}\\
	  \Rightarrow Lw &= D^t\Lambda \tilde{G},\\
\end{split}
\end{equation*}
which is an equation of the form $Ax = b$, where $x=w$ is an unknown, and we can compute $L$, $D$, $\Lambda$ and $\tilde{G}$. $A$ is a symmetric, positive definite matrix. We solve this equation for $w$ which is the geodesic distance function we want up to an additive constant. We shift this distance function so that the smallest distance is $0$. This gives us our final geodesic distances.

\section{Evaluation and Results}
We compare our approximate geodesics using wave propagation against geodesics from heat \cite{Crane_et_al.2013}, as well as against exact geodesics, described in \cite{Surazhsky_et_al.2005} and implemented by Kirsanov [Kirsanov et al] (Table~\ref{table:error}). We compute two errors to assess the accuracy of our method:
\begin{itemize}
\item Raw error: the absolute difference between geodesics computed using our method and the exact method.
\item Relative error: the ratio between the raw error and the exact geodesic distance. 
\end{itemize}

\begin{figure*}[!t]
	\centering
\makebox[46pt][r]{\textbf{Exact}}
	\begin{subfigure}{33mm} 
		\includegraphics[scale=0.20]{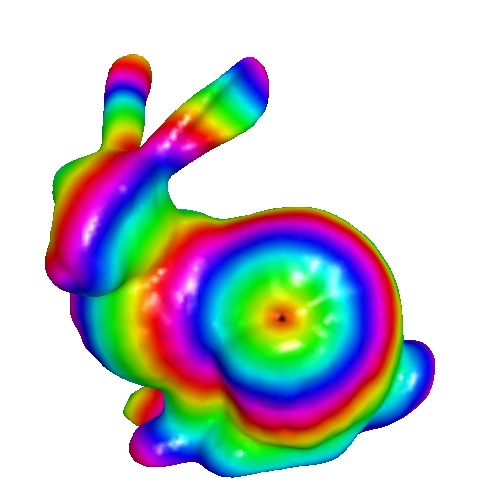}
	\end{subfigure}
	\begin{subfigure}{30mm}
		\includegraphics[scale=0.215]{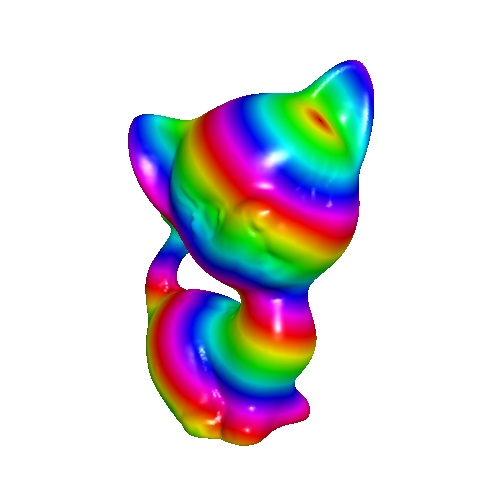}
	\end{subfigure}
	\begin{subfigure}{33mm}
		\includegraphics[scale=0.23]{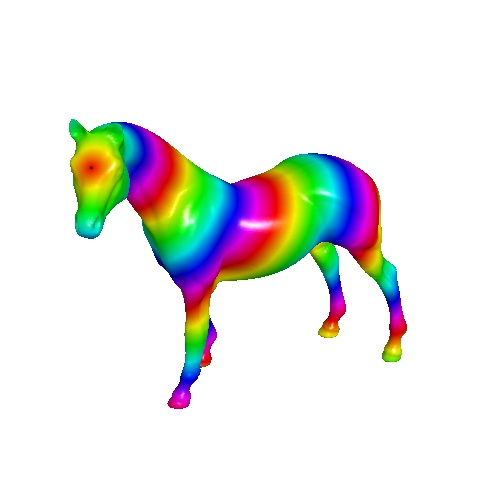}
	\end{subfigure}
	\begin{subfigure}{32mm}
		\includegraphics[scale=0.215]{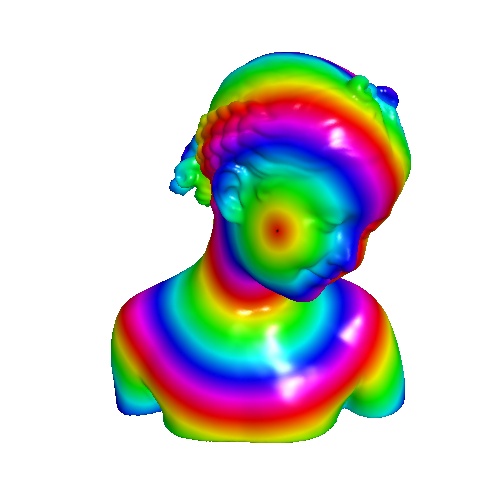}
	\end{subfigure}
	\begin{subfigure}{30mm}
		\includegraphics[scale=0.21]{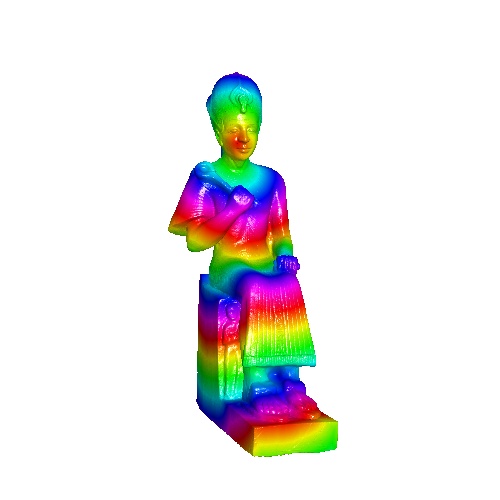}
	\end{subfigure}\\ \vspace{-6mm}
\makebox[46pt][r]{\textbf{Wave}}
	\begin{subfigure}{33mm}
		\includegraphics[scale=0.20]{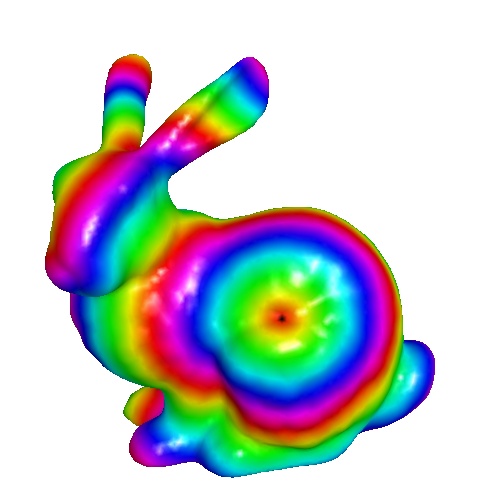}
	\end{subfigure}
	\begin{subfigure}{30mm}
		\includegraphics[scale=0.215]{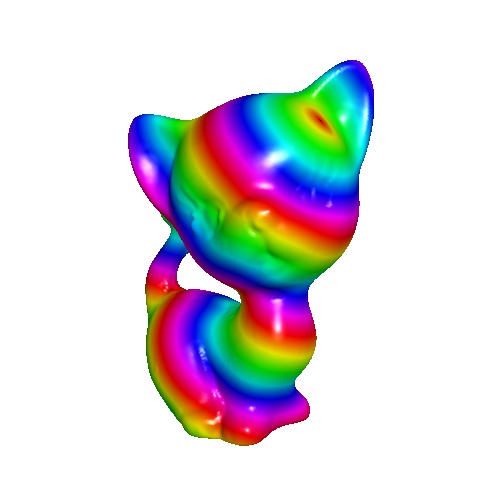}
	\end{subfigure}
	\begin{subfigure}{33mm}
		\includegraphics[scale=0.23]{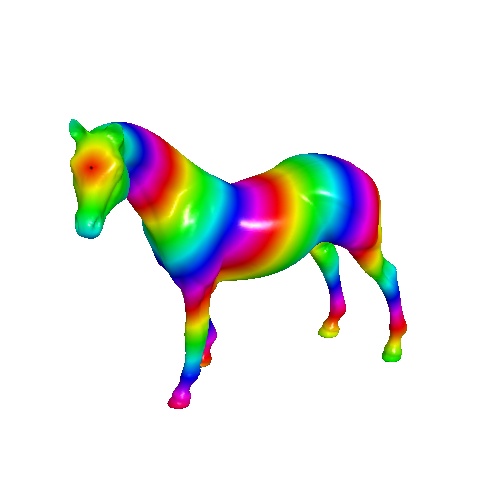}
	\end{subfigure}
	\begin{subfigure}{32mm}
		\includegraphics[scale=0.215]{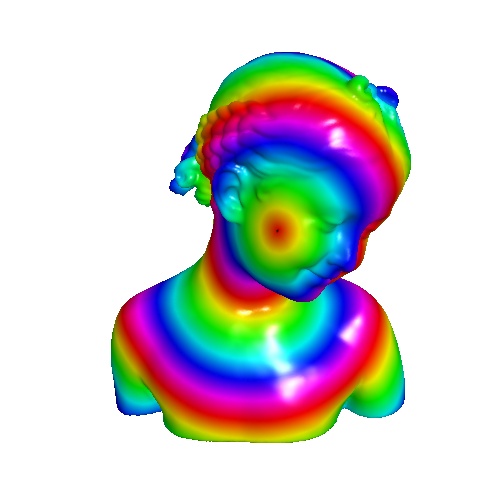}
	\end{subfigure}
	\begin{subfigure}{30mm}
		\includegraphics[scale=0.21]{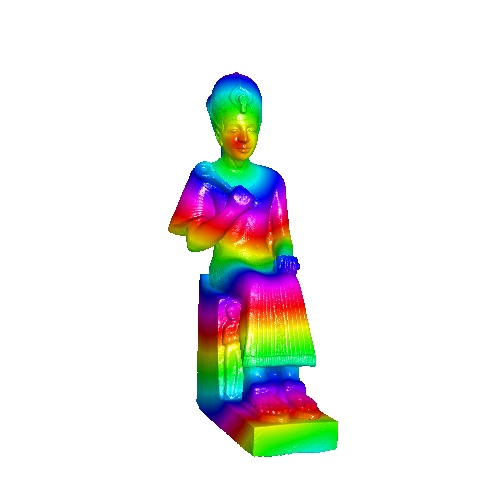}
	\end{subfigure}\\ \vspace{-6mm}
\makebox[46pt][r]{\textbf{Raw Error}}
	\begin{subfigure}{33mm}
		\includegraphics[scale=0.20]{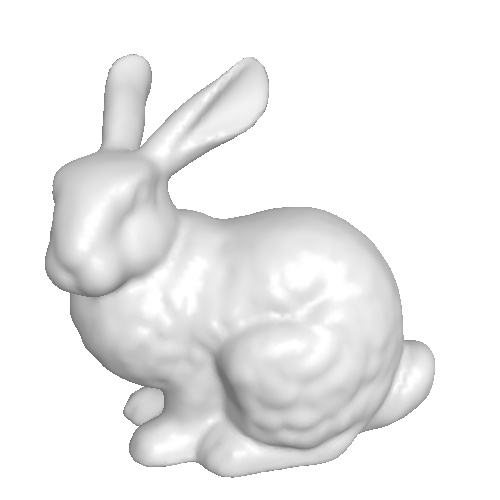}
	\end{subfigure}
	\begin{subfigure}{30mm}
		\includegraphics[scale=0.215]{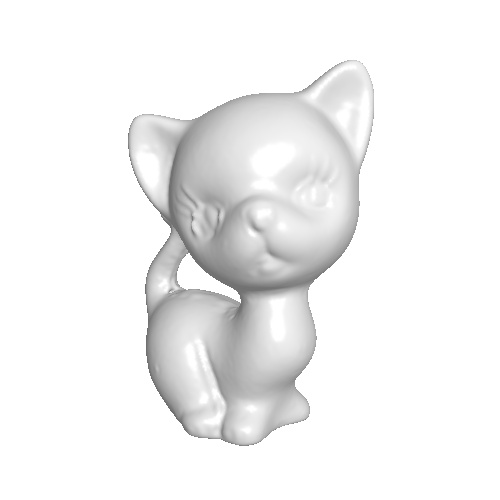}
	\end{subfigure}
	\begin{subfigure}{33mm}
		\includegraphics[scale=0.23]{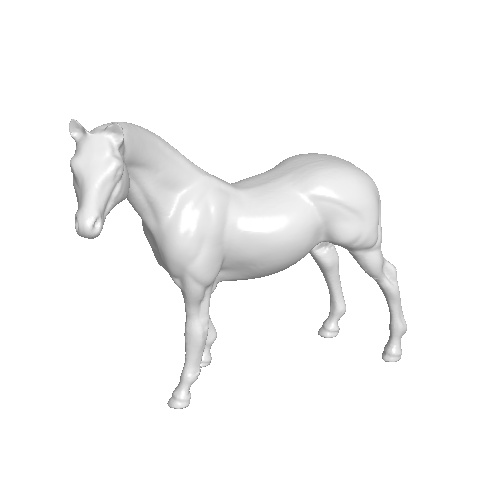}
	\end{subfigure}
	\begin{subfigure}{32mm}
		\includegraphics[scale=0.215]{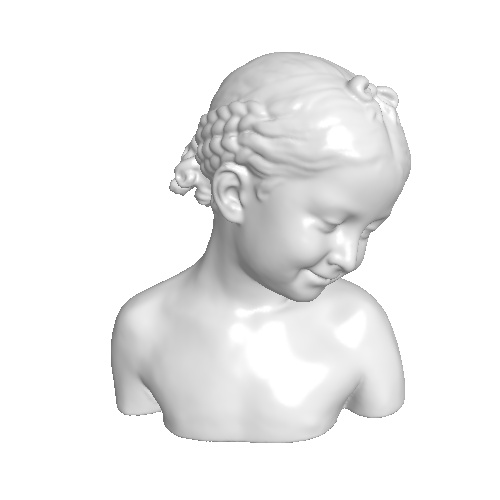}
	\end{subfigure}
	\begin{subfigure}{30mm}
		\includegraphics[scale=0.21]{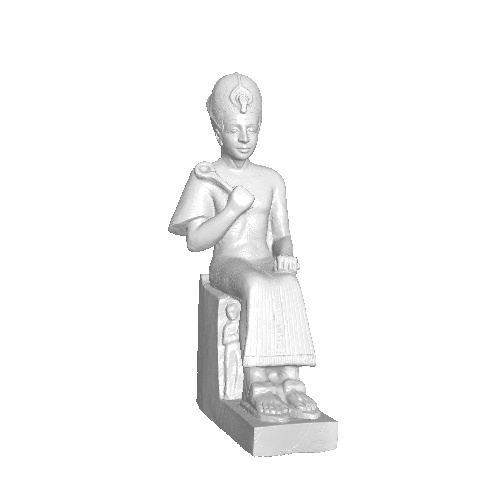}
	\end{subfigure}\\ \vspace{-5mm}
\makebox[46pt][r]{\textbf{Raw Error}}
	\begin{subfigure}{32mm}
		\includegraphics[scale=0.21]{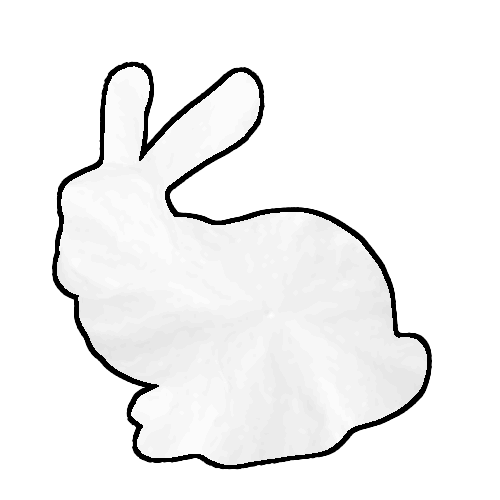}
	\end{subfigure}
	\begin{subfigure}{33mm}
		\includegraphics[scale=0.23]{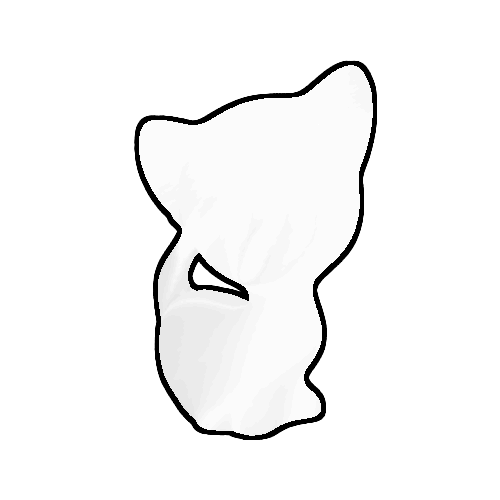}
	\end{subfigure}
	\begin{subfigure}{34mm}
		\includegraphics[scale=0.20]{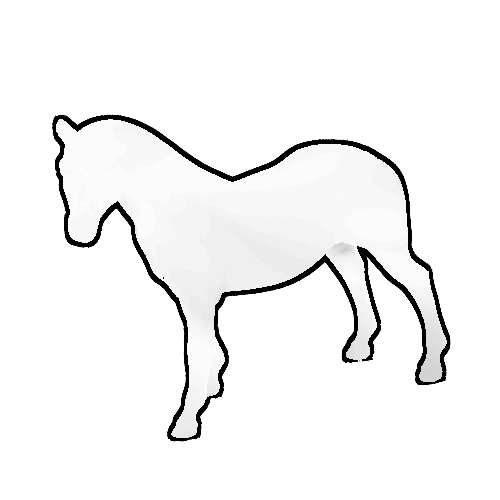}
	\end{subfigure}
	\begin{subfigure}{37mm}
		\includegraphics[scale=0.20]{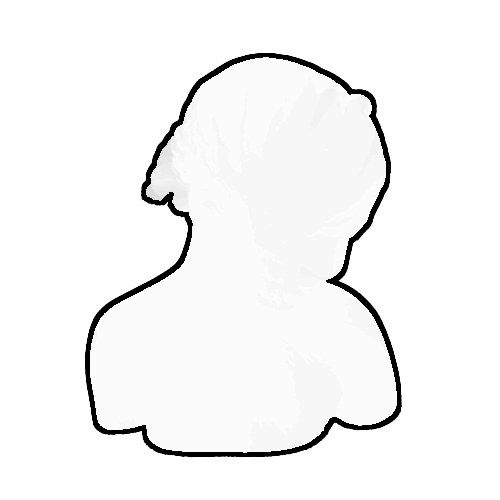}
	\end{subfigure}
	\begin{subfigure}{21mm}
		\includegraphics[scale=0.15]{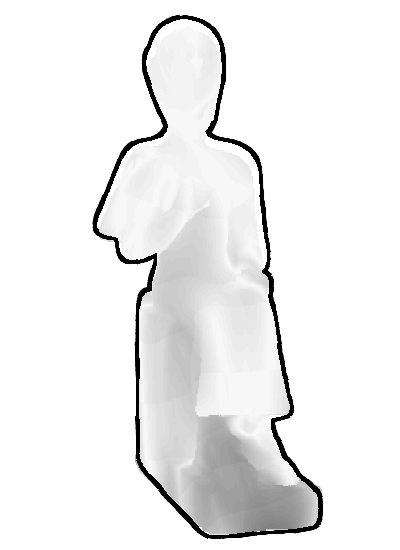}
	\end{subfigure}\\ \vspace{-6mm}
\makebox[46pt][r]{\textbf{Relative Error}}
	\begin{subfigure}{33mm}
		\includegraphics[scale=0.20]{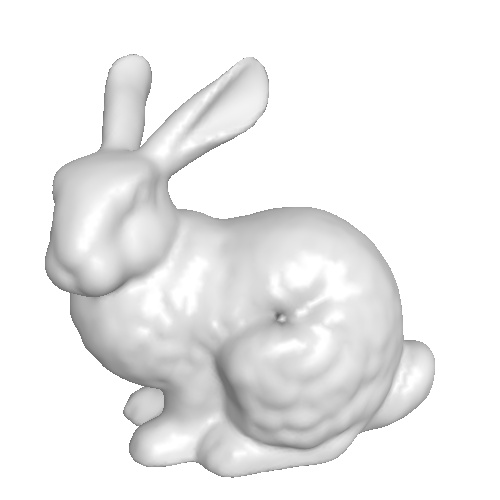}
	\end{subfigure}
	\begin{subfigure}{30mm}
		\includegraphics[scale=0.215]{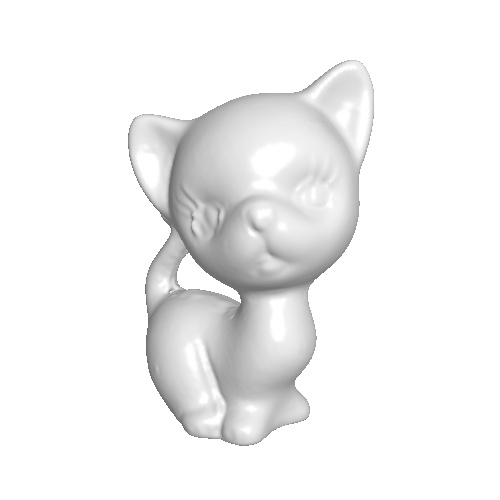}
	\end{subfigure}
	\begin{subfigure}{33mm}
		\includegraphics[scale=0.23]{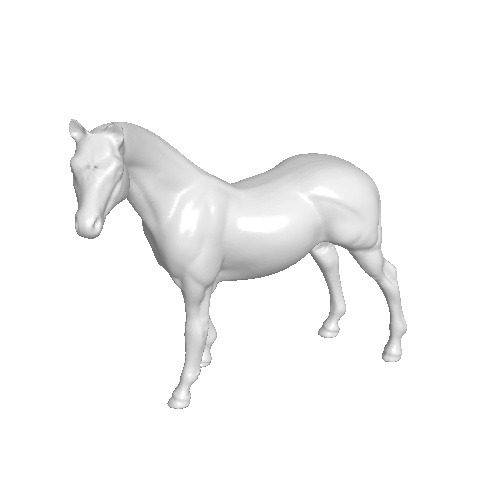}
	\end{subfigure}
	\begin{subfigure}{32mm}
		\includegraphics[scale=0.215]{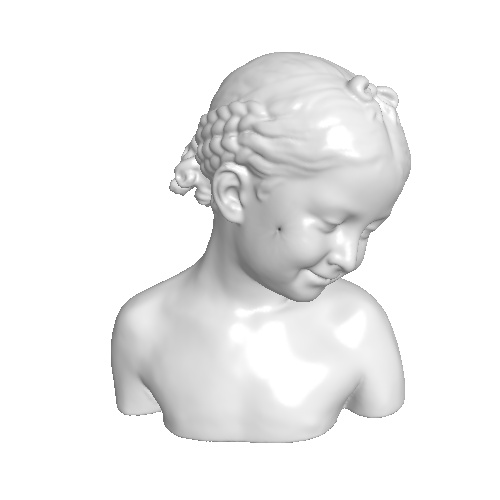}
	\end{subfigure}
	\begin{subfigure}{30mm}
		\includegraphics[scale=0.21]{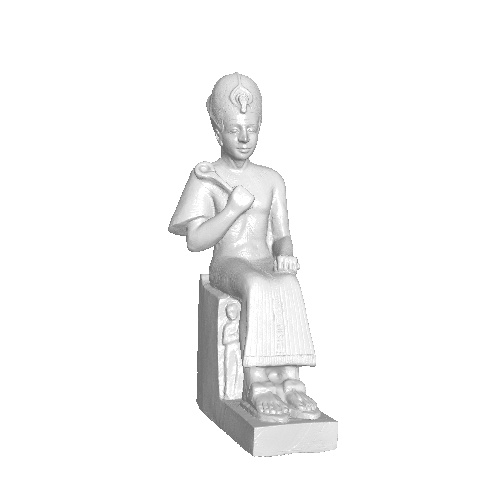}
	\end{subfigure}\\ \vspace{-5mm}
\makebox[46pt][r]{\textbf{Relative Error}}
	\begin{subfigure}{32mm}
		\includegraphics[scale=0.21]{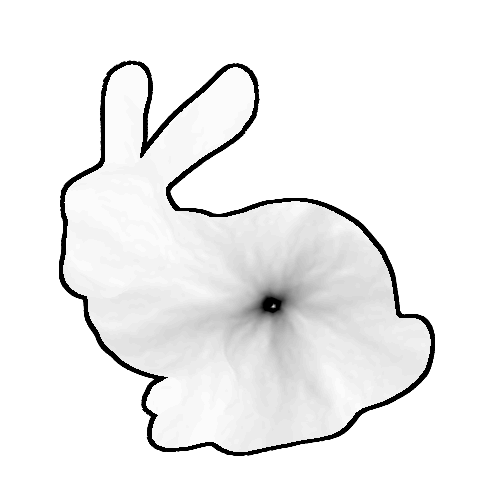}
	\end{subfigure}
	\begin{subfigure}{33mm}
		\includegraphics[scale=0.23]{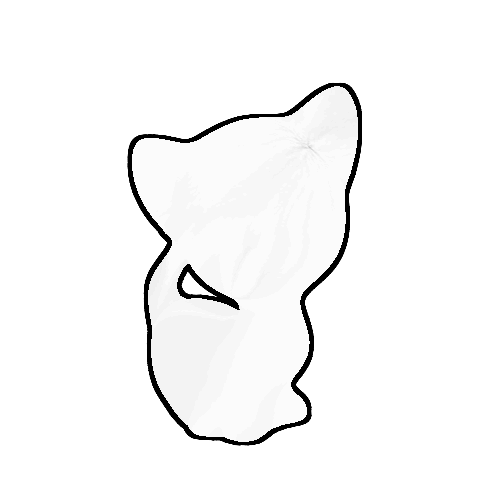}
	\end{subfigure}
	\begin{subfigure}{34mm}
		\includegraphics[scale=0.20]{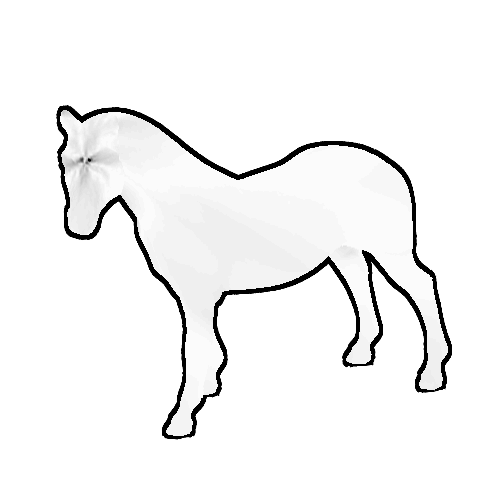}
	\end{subfigure}
	\begin{subfigure}{37mm}
		\includegraphics[scale=0.20]{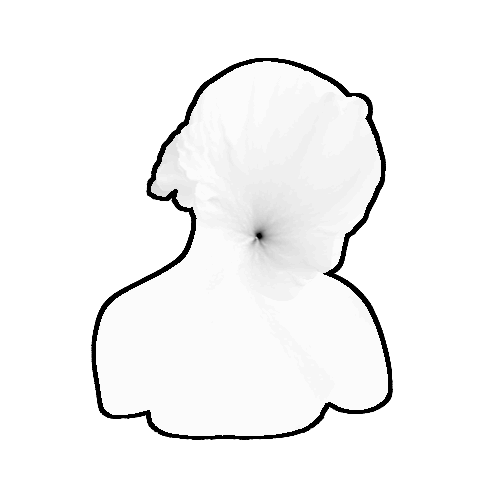}
	\end{subfigure}
	\begin{subfigure}{21mm}
		\includegraphics[scale=0.15]{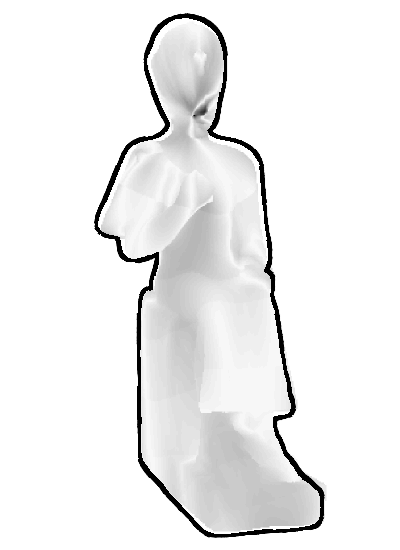}
	\end{subfigure}\\ 
	\caption{Comparison of the wave method with the exact geodesic method on (left to right) the Bunny, Kitten, Horse, Bimba, and Ramesses models. This figure shows isocontours from exact geodesics (top row), isocontours from geodesics obtained using our method (2nd row), raw error (3rd row), raw error without lighting (4th row), relative error (5th row), and relative error without lighting (last row). The images showing error without lighting are rescaled to show where the errors are located. The black dot in the 1st and 2nd rows indicates the source.}
	\label{fig:compare}
\end{figure*}

These metrics are computed at each vertex and then averaged. We also visually compare our method with the exact method in Fig.~\ref{fig:compare}. We indicate zero error by an RGB value of $1$, and an error of $1$ by an RGB value of $0$ (Fig.~\ref{fig:compare}, rows $3$, $5$), since all our errors are smaller than $1$ (Table~\ref{table:error}). However, these errors are hard to visualize because they are much smaller than $1$, and, in addition, shadows due to the lighting model also manifest themselves in similar colors as the errors. Therefore, we visualize the models without lighting, and rescale the intensity values so that values between $0.7$ and $1$ are remapped to values between $0$ and $1$ (Fig.~\ref{fig:compare}, rows $4$, $6$). We see that raw error is higher away from the source, since the wave becomes more smoothed as it travels away from the source. Some more results using our method can be seen in Fig.~\ref{fig:more}.

\begin{figure}[!b]
  \centering
  \includegraphics[width=0.5\textwidth]{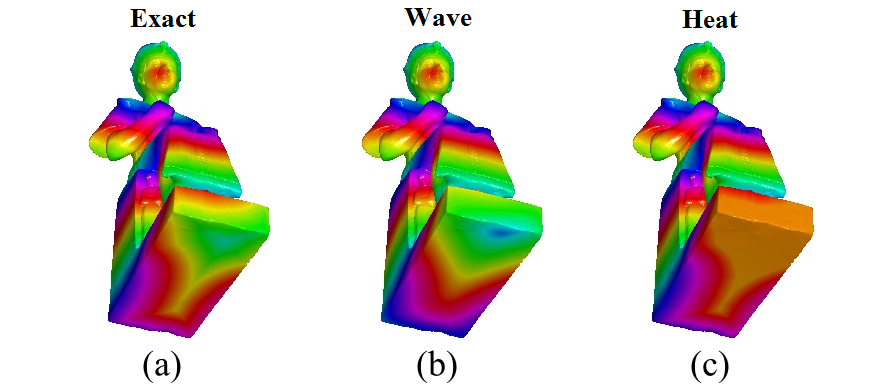}
  \caption{The Ramesses model shows that our method can handle very large meshes (1.6 million faces), and approximate geodesics reliably even far away from the source. The isocontours on the base of the Ramesses model plotted using geodesics from the wave method (b) show a more similar coloring compared with exact geodesics (a), than those plotted using heat geodesics (c). However, the shape of the isocontours using our method (b) is more smoothed than both the exact method (a) and the heat method (c).}
  \label{fig:away}
\end{figure}

\begin{figure*}[!t]
  \centering
  \includegraphics[width=1\textwidth]{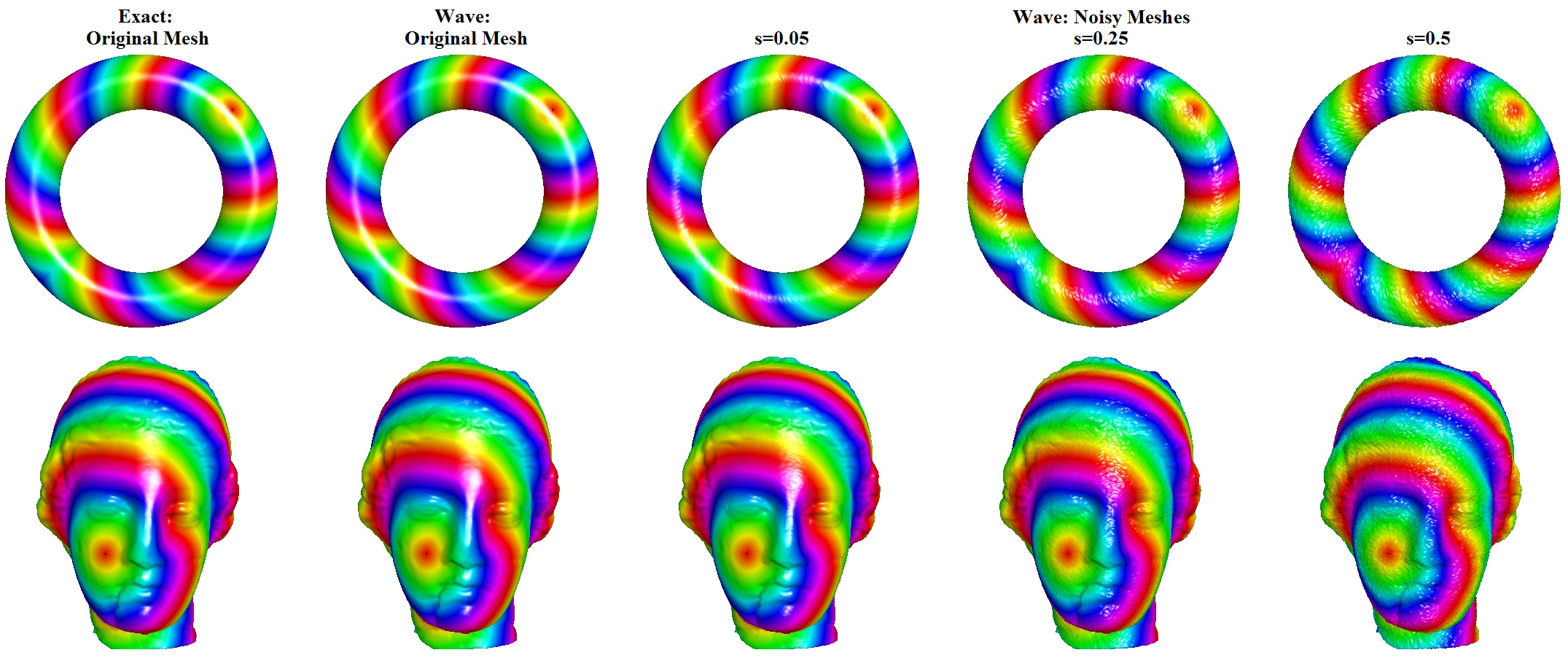}
  \caption{Geodesic distance computed on the torus (top) and Igea (bottom) models. For both shapes, (from left to right) we have the exact geodesics and geodesics from waves on the original models, as well as geodesics from waves on the models with added noise, where s = 0.05, 0.25 and 0.5.}
  \label{fig:noise}
\end{figure*}

\begin{figure*}[t]
\vspace{25mm}
\centering
  	\begin{subfigure}[t]{0.5\textwidth}
  		\includegraphics[width=\textwidth]{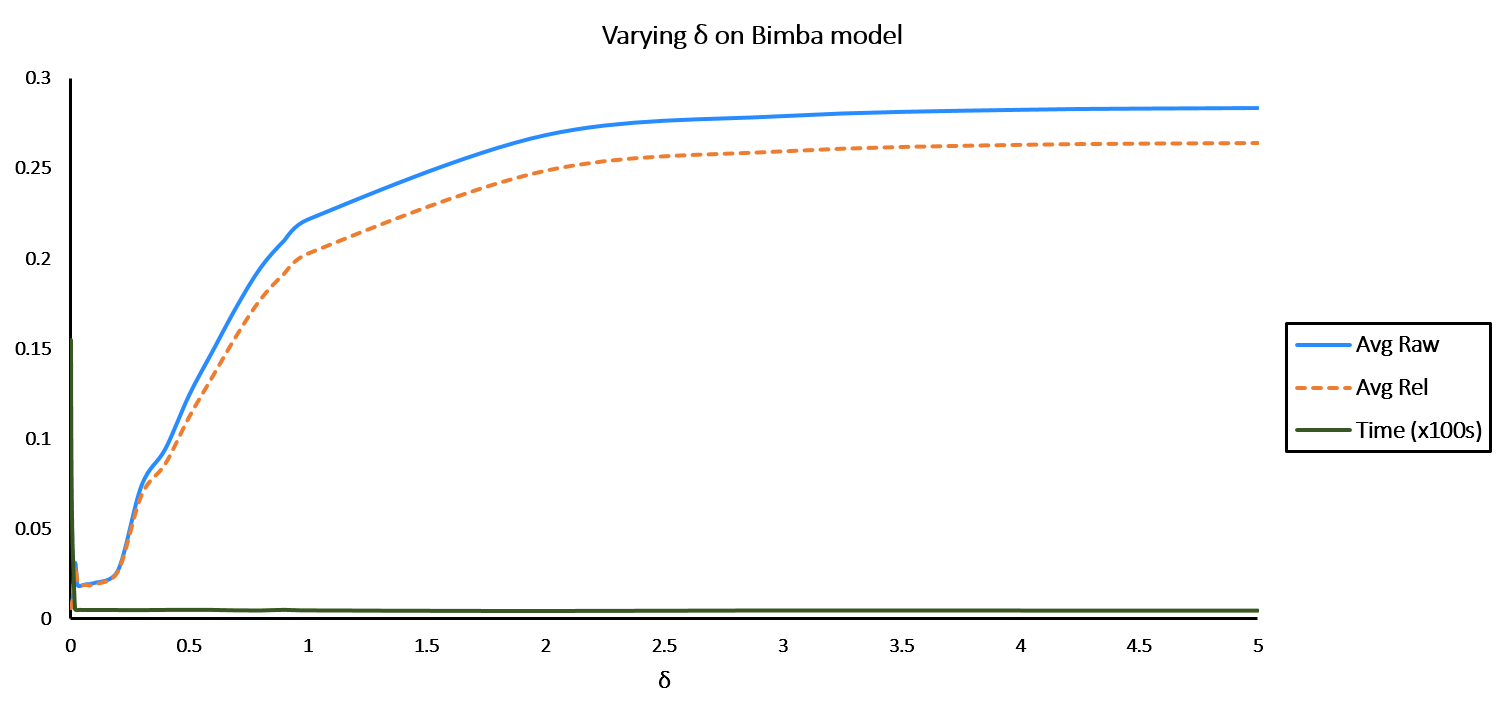}
  		\caption{}
  		\label{fig:delta}
	\end{subfigure}%
	\begin{subfigure}[t]{0.5\textwidth}
 		 \includegraphics[width=\textwidth]{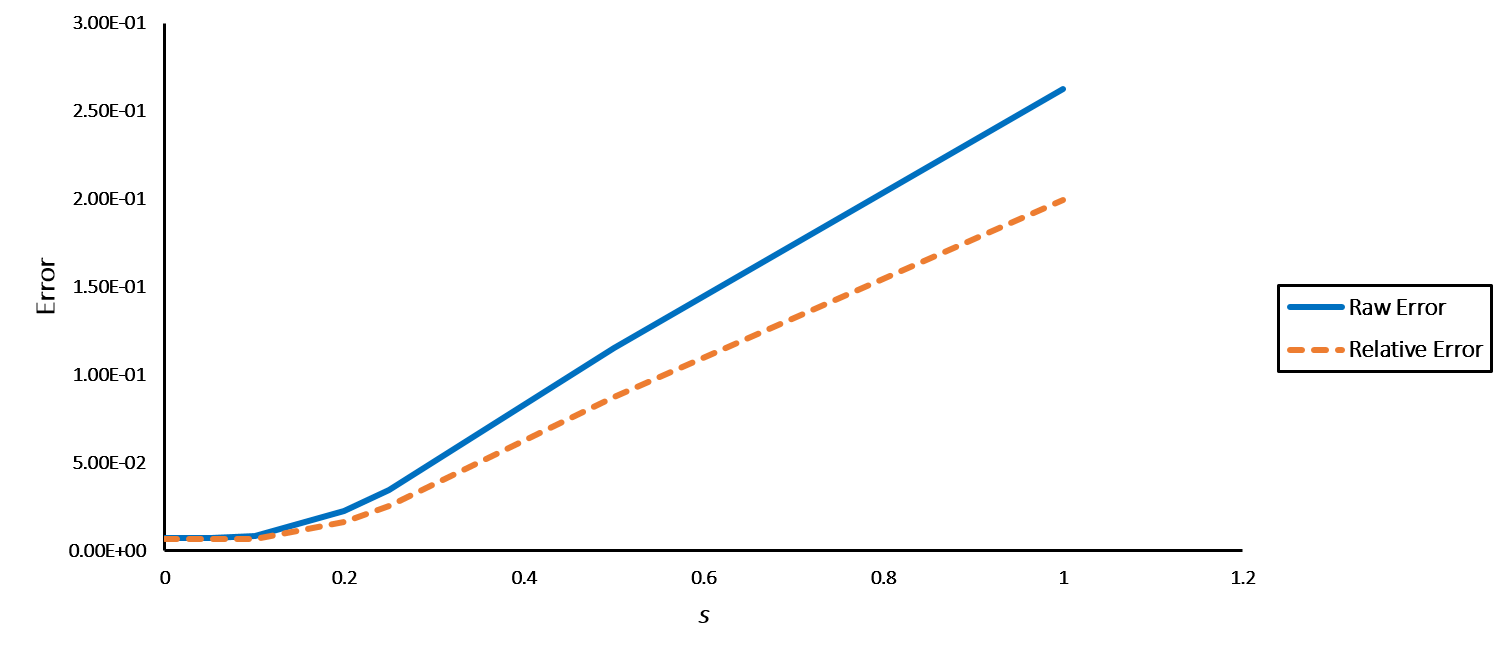}
 		\caption{}
		\label{fig:noise_error}
	\end{subfigure}
\caption{(a) This graph shows how the time-step, $\delta$, affects raw and relative errors, and time ($\times100$s) taken to compute geodesics on the bimba model. As $\delta$ increases, errors also increase, while the time taken decays very quickly. For values of $\delta$ between $0.005$ and $0.1$, the errors are small, and the time taken to compute geodesics is under $10$s; (b) Errors computed for increasing noise.}
\end{figure*}

Both our method and the heat method solve two linear systems. The heat method requires that both these systems be solved only once, whereas our method requires us to solve one of the two linear systems repeatedly until our propagating wave has traveled across the shape. This makes our method slower than the heat method. However, our method can be sped up by changing the parameters that increase the speed of the wave traveling on the surface of our mesh. The runtime of the wave method is affected by parameter $\delta$. The larger these parameters are, the faster the wave propagates. However, errors in geodesics approximated using our method increase as the time-step, or $\delta$, increases (Fig.~\ref{fig:delta}). The potential for aliasing also increases as $\delta$ gets larger. On the other hand, increasing the resolution of a mesh increases the accuracy of our method, while also increasing the time required to compute the geodesics. However, we found that our method scales better than the exact method (Table.~\ref{table:resolution}).


We further evaluated the robustness of our method in the presence of noise~\cite{Szymon2004}. We added noise to each vertex by shifting each vertex by \[O(s\times l),\] where $l$ is the median edge length of all the edges incident on vertex $v$, and $s$ varies between $0$ and $1$. Our method remained fairly robust in the presence of noise (Fig.~\ref{fig:noise}), producing good approximations even in the presence of considerable amounts of noise (Fig.~\ref{fig:noise_error}). We also compared errors on smoothed as well as sharpened versions of models~\cite{Szymon2004}. We smoothed our models by performing $m$ iterations of simple umbrella smoothing \cite{Kobbelt_et_al.1998}, where $m$ ranges between $10$ and $50$. In order to sharpen our models, we first smooth our model using a smoothing kernel with radius equal to $\sigma$, which is given by 
\begin{equation*}
	\sigma = s \times l,
\end{equation*}
where $l$ is the median edge length of all the edges incident on vertex $v$. In our experiments, $s$ varies between $0.005$ and $5$. Once the position of each vertex after smoothing is obtained, we add  to the original vertices twice the difference between the original and the new vertices to obtain a sharpened mesh. Our method is quite robust to smoothening and sharpening, with minimal changes in error. As meshes become sharper, the errors tend to increase slightly, whereas the errors decrease as meshes become smoother. The increase in error with increasing sharpness arises from discretization. 

\begin{figure*}[t]
\vspace{2.9mm}
  \centering
  \includegraphics[width=1\textwidth]{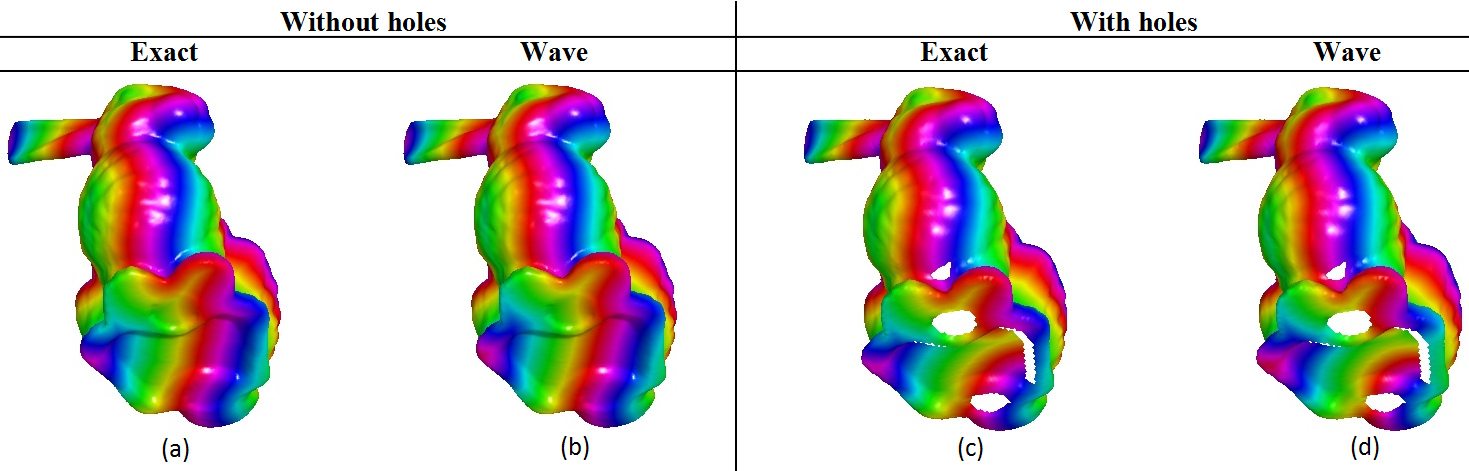}
  \caption{(a) Exact geodesics, and (b) approximate geodesics obtained using the wave method, plotted on the Stanford bunny model without holes. (c) Exact geodesics, and (d) geodesics using waves, plotted on the bunny model with holes.}
  \label{fig:bunny}
\end{figure*}

Our method also performs well in the presence of holes. The geodesics computed on a mesh without holes, and the same mesh with holes, are comparable, and both have small errors. For instance, on the bunny model without holes (Fig.~\ref{fig:bunny}a,b), our approximations had a raw error of $0.0057$, and a relative error of $0.005$. The errors on the bunny model with holes (Fig.~\ref{fig:bunny}c,d) was slightly higher. The raw error for the model with holes was $0.013$, and the relative error was $0.012$.

\section{Conclusion and Future Work}

We present a new method to approximate geodesic distances on shapes. Our method produces reliable approximations to exact geodesic distances. We also show that our method scales well with the size of the mesh. Our method can therefore be used in several applications that require approximate geodesics. However, there are several areas that can still be explored. For instance, the method for tracking wave propagation can be improved, and the current height function for $\epsilon$ can also be further optimized. Our method is also very sensitive to the value of $\delta$. Making it robust to $\delta$ is an important future direction of work. 

However, most importantly, our current divergence computation can be improved by storing the gradient of our pseudo-distance function at triangle faces, rather than storing the mean gradient at every edge (Eq.~\ref{eq:edge_grad}). Although we have not implemented this, the details of this implementation are explained in Appendix~\ref{apdx:A}. 

This method has several applications. One application, for instance, is estimating Voronoi regions by propagating waves from multiple sources. Another application is that it can be used in physically based rendering of subsurface scattering to quickly estimate points that are geodesically close by, instead of computing nearby points using Euclidean distances.

\section*{Acknowledgments}

This research was supported by the Johns Hopkins University internal funds. We would also like to thank the Stanford Computer Graphics Laboratory and AIM@Shape for providing us with meshes to test our method. 

\bibliographystyle{acmsiggraph}
\bibliography{wavebib}

\begin{thebibliography}{\protect\citename{Surazhsky et~al\mbox{.} }2005}

\bibitem[\protect\citename{Chen et~al\mbox{.} }2008]{Chen_et_al.2008}
{\sc Chen, Y., Davis, T., Hager, W., and Rajamanickam, S.}
\newblock 2008.
\newblock Algorithm 887: Cholmod, supernodal sparse cholesky factorization and
  update/downdate.
\newblock {\em ACM Transactions on Mathematical Software (TOMS) 35}, 3,
  22:1--22:14.

\bibitem[\protect\citename{Crane et~al\mbox{.} }2013]{Crane_et_al.2013}
{\sc Crane, K., Weischedel, C., and M, W.}
\newblock 2013.
\newblock Geodesics in heat: A new approach to computing distance based on heat
  flow.
\newblock {\em ACM Transactions on Graphics (TOG) 32}, 5, 152:1--152:11.

\bibitem[\protect\citename{Hysing and Turek }2005]{Hysing_and_Turek2005}
{\sc Hysing, S., and Turek, S.}
\newblock 2005.
\newblock The eikonal equation: Numerical efficiency vs. algorithmic complexity
  on quadrilateral grids.
\newblock {\em Proceedings of ALGORITMY\/}, 22--31.

\bibitem[\protect\citename{Kimmel and Sethian }1998]{Kimmel_and_Sethian1998}
{\sc Kimmel, R., and Sethian, J.}
\newblock 1998.
\newblock Computing geodesic paths on manifolds.
\newblock In {\em Proceedings of the National Academy of Sciences, USA},
  vol.~95, 8431--8435.

\bibitem[\protect\citename{Kobbelt et~al\mbox{.} }1998]{Kobbelt_et_al.1998}
{\sc Kobbelt, L., Campagna, S., Vorsatz, J., and Seidel, H.}
\newblock 1998.
\newblock Interactive multi-resolution modeling on arbitrary meshes.
\newblock {\em Proceedings of the 25th Annual Conference on Computer Graphics
  and Interactive Techniques\/}, 105--114.

\bibitem[\protect\citename{Mitchell et~al\mbox{.} }1987]{Mitchell_et_al.1987}
{\sc Mitchell, J., Mount, D., and Papadimitriou, C.}
\newblock 1987.
\newblock {The Discrete Geodesic Problem}.
\newblock {\em SIAM Journal on Computing 16}, 4, 647--668.

\bibitem[\protect\citename{Rusinkiewicz }2004]{Szymon2004}
{\sc Rusinkiewicz, S.}
\newblock 2004.
\newblock trimesh2.

\bibitem[\protect\citename{Surazhsky et~al\mbox{.} }2005]{Surazhsky_et_al.2005}
{\sc Surazhsky, V., Surazhsky, T., Kirsanov, D., Gortler, S., and Hoppe, H.}
\newblock 2005.
\newblock Fast exact and approximate geodesics on meshes.
\newblock {\em ACM Transactions on Graphics (TOG) 24}, 3, 553--560.

\end{thebibliography}
\newpage

\begin{figure*}[t]
  \centering \vspace{-3mm}
  \includegraphics[width=1\textwidth]{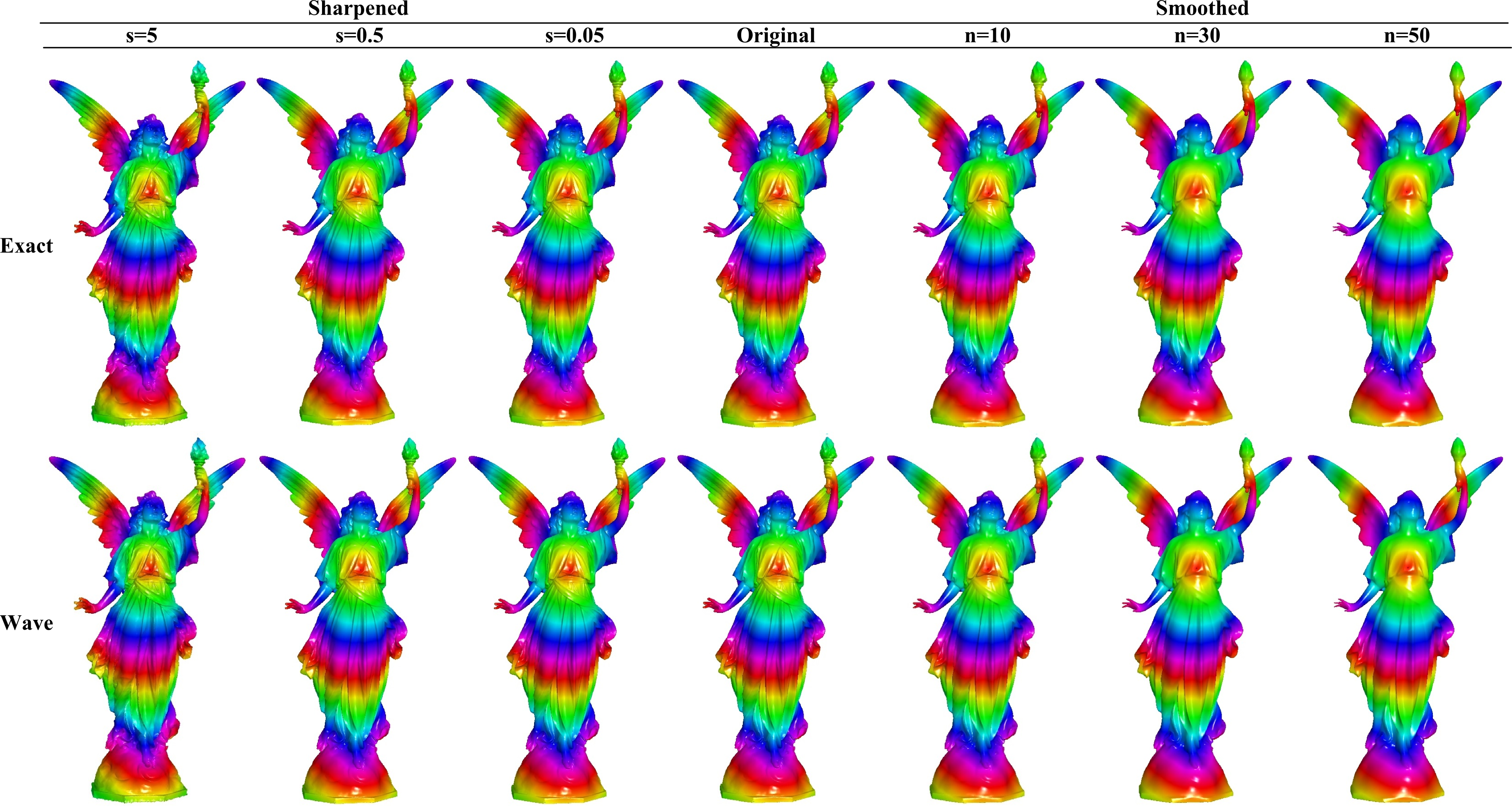}
  \caption{Geodesic distances computed on the Lucy model. The top row shows the exact geodesics, while the bottom row shows geodesics from waves. From left to right, we have the Lucy model sharpened with $s = 5, 0.5,$ and $0.05$, where $\sigma = s * l$, the original lucy model, and the lucy model smoothed with $n = 10, 30,$ and $50$ iterations of simple umbrella smoothing.}
  \label{fig:sharpsmooth}
\end{figure*}

\begin{figure*}[b]
\centering
  	\begin{subfigure}[t]{0.5\textwidth}
	\centering
  		\includegraphics[width=\textwidth]{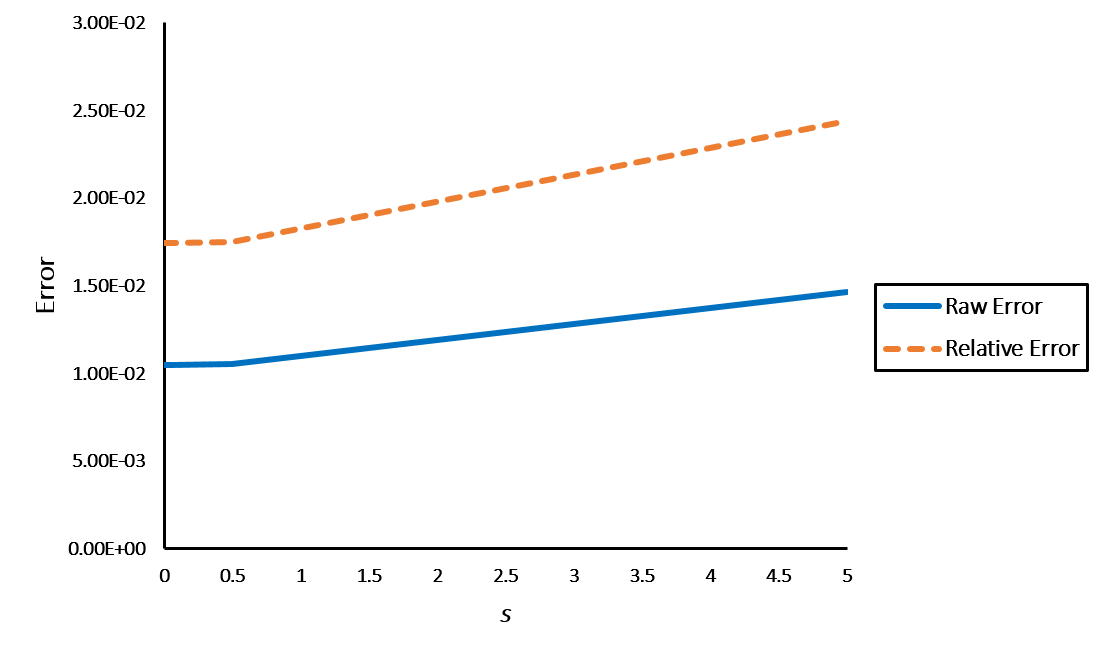}
  		\caption{}
  		\label{fig:sharp_error}
	\end{subfigure}%
	\begin{subfigure}[t]{0.5\textwidth}
	\centering
  		\includegraphics[width=\textwidth]{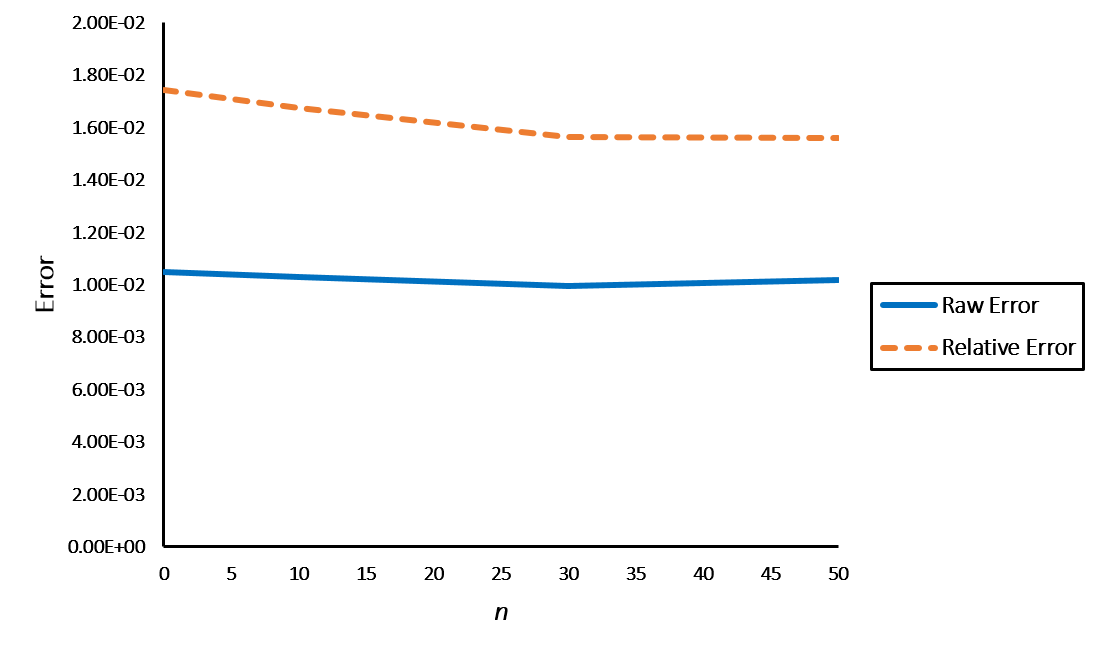}
 		\caption{}
  		\label{fig:smooth_error}
	\end{subfigure}
\caption{(a) Errors computed for increasing sharpness, and (b) errors computed for increasing smoothness.}
\end{figure*}

\clearpage

\appendix{APPENDIX}

\section{An alternative formulation for the computation of geodesics}
\label{apdx:A}
In Eq.~\ref{eq:edge_grad}, we compute, for every edge, the mean of the gradients at the two triangles incident upon the edge, and save this mean gradient at the edge. This, however, leads to smoothing of gradients, and therefore, loss of information, which explains why the errors in Fig.~\ref{fig:sharp_error} go up as the sharpness in the model increases. 

Instead, we can save the normalized gradient, $\bar{G}$, at each triangle face, and compute the divergence of this gradient using the following equation: 
\begin{equation*}
\begin{split}
	  Div_i = \int \langle \nabla h_i, \bar{G} \rangle,
\end{split}
\end{equation*}
where $h_i$ is the hat-basis function at vertex $v_i$, and $\bar{G}$ is the normalized gradient per mesh face. Divergence can be discretized as:
\begin{equation*}
\begin{split}
	  Div_i(\bar{G}) = \sum_{f\in N(v_i)} \langle \nabla h_i\big|_f . \bar{G}\big|_f \rangle . |f|,
\end{split}
\end{equation*}
where the gradient of $h_i$ is restricted to faces $f$ that are incident upon the vertex $v_i$ since the gradients of the hat-basis functions and the vector field, $\bar{G}$, are constant per face, and the gradient of the hat-basis functions is supported in the one-ring. $|f|$ is the area of triangle face $f$.

Similarly, the Laplacian for every vertex pair $(i, j)$ can be computed as
\begin{equation*}
\begin{split}
	  L_{ij} = \int \langle \nabla h_i , \nabla h_j  \rangle, 
\end{split}
\end{equation*}
which can be discretized as
\begin{equation*}
\begin{split}
	  L_{ij} = \sum_{f\in N(v_i) \cap N(v_j)} \langle \nabla h_i \big|_{f} .  \nabla h_j\big|_{f}\rangle |f|.
\end{split}
\end{equation*}
Since the Laplacian is the divergence of the gradient, we can solve the equation 
\begin{equation*}
\begin{split}
	 \Delta w =  Div_i(\bar{G})
\end{split}
\end{equation*}
to compute the underlying function, $w$, that approximates the geodesic distance function.

\begin{figure}[t]
	\centering
	\vspace{-67.5mm}
	\begin{subfigure}{33mm} 
		\includegraphics[scale=0.215]{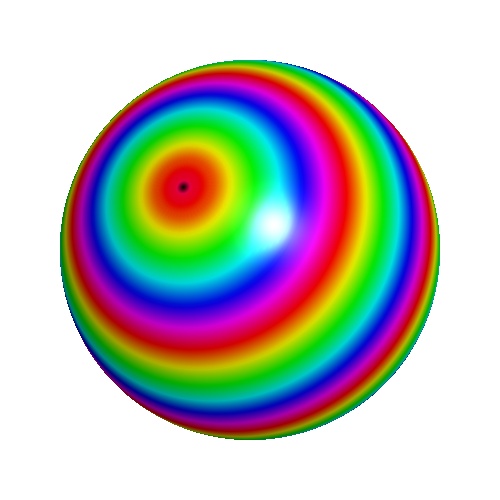}
	\end{subfigure}
	\begin{subfigure}{30mm}
		\includegraphics[scale=0.22]{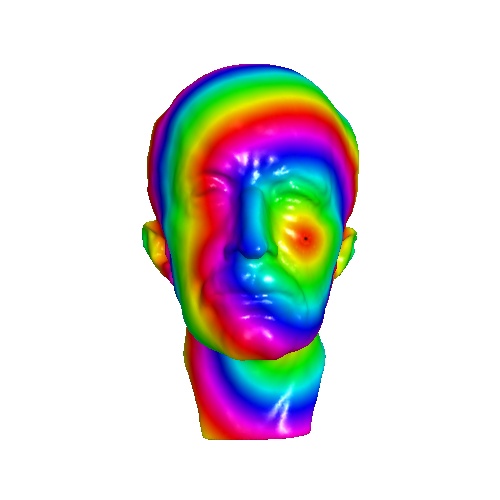}
	\end{subfigure}
	\begin{subfigure}{33mm}
		\includegraphics[scale=0.22]{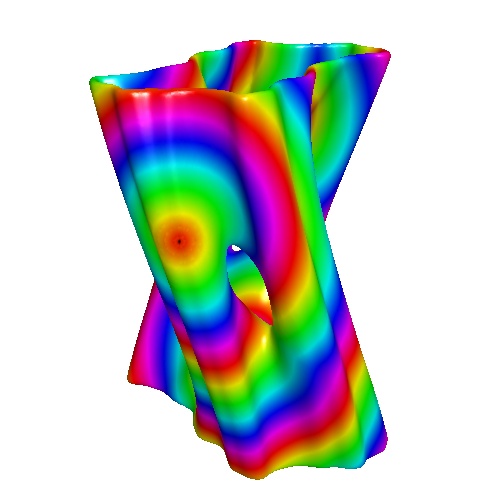}
	\end{subfigure}
	\begin{subfigure}{32mm}
		\includegraphics[scale=0.22]{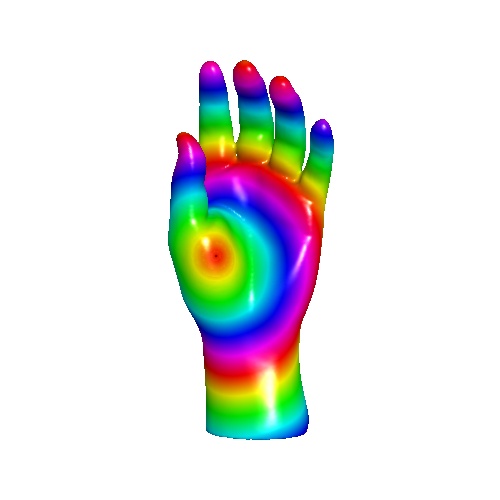}
	\end{subfigure}
	\begin{subfigure}{33mm}
		\includegraphics[scale=0.22]{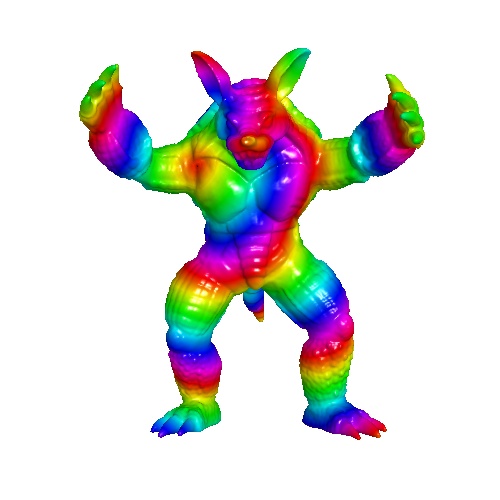}
	\end{subfigure}
	\begin{subfigure}{32mm}
		\includegraphics[scale=0.22]{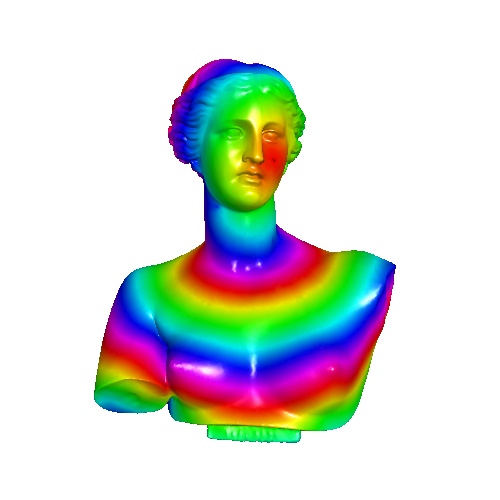}
	\end{subfigure}
	\begin{subfigure}{30mm}
		\includegraphics[scale=0.22]{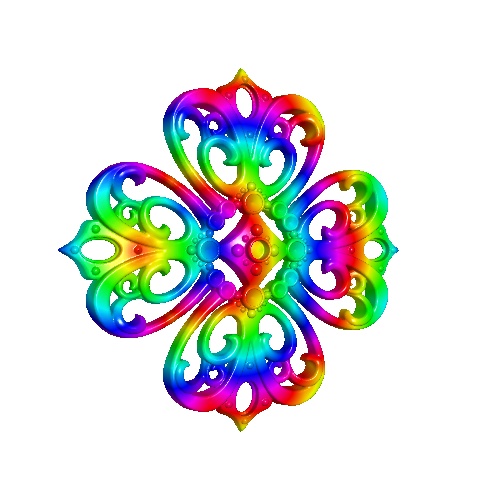}
	\end{subfigure}
	\begin{subfigure}{30mm}
		\includegraphics[scale=0.22]{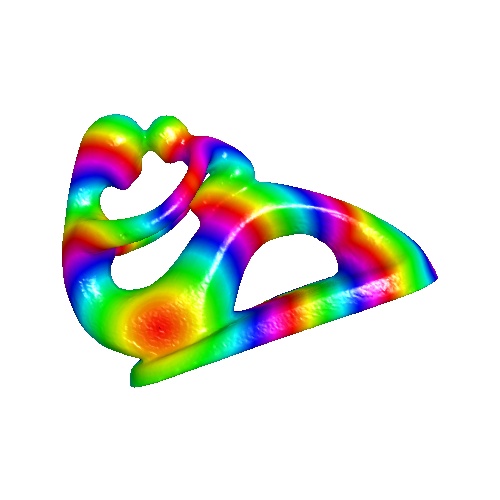}
	\end{subfigure}
	\caption{Geodesics computed using our method on more models: (from left to right, top to bottom) Sphere, Max Planck, Julia vase, Hand, Armadillo man, Aphrodite, Filigree and Fertility models.}
	\label{fig:more}
\end{figure}

\clearpage

\end{document}